# Title: Holistic Multi-scale Imaging of Oxygen Reduction Reaction Catalyst Degradation in Operational Fuel Cells

**Authors:** Isaac Martens[1,2], Antonis Vamvakeros[1,3,4], Nicolas Martinez[5], Raphaël Chattot[1], Janne Pusa[1], Maria Valeria Blanco[1], Elizabeth A. Fisher[2], Tristan Asset[6], Sylvie Escribano[5], Fabrice Micoud[5], Tim Starr[7], Alan Coelho[8], Veijo Honkimäki[1], Dan Bizzotto[2], David P. Wilkinson[9], Simon D.M. Jacques[4], Frédéric Maillard[6], Laetitia Dubau[6], Sandrine Lyonnard[5*], Arnaud Morin[10*], Jakub Drnec[1*]

**Affiliations:**

[1]European Synchrotron Radiation Facility, Grenoble, France.

[2]Advanced Materials and Process Engineering Laboratory, Department of Chemistry, University of British Columbia, Vancouver, Canada.

[3]Department of Chemistry, University College London, London, UK.

[4]Finden Ltd., Oxfordshire, UK.

[5]Univ. Grenoble Alpes, CEA, CNRS, IRIG, SyMMES, F-38054 Grenoble, France.

[6]Univ. Grenoble Alpes, Univ. Savoie Mont Blanc, CNRS, Grenoble INP, LEPMI, 38000, Grenoble, France.

[7]Independent researcher, St. Johann im Tirol, Austria.

[8]Coelho Software, Brisbane, Australia.

[9]Department of Chemical and Biological Engineering, University of British Columbia, Vancouver, Canada.

[10]Université Grenoble Alpes, CEA, 17 Avenue des Martyrs, F-38000 Grenoble.

*Correspondence to: sandrine.lyonnard@cea.fr, arnaud.morin@cea.fr, drnec@esrf.fr

**Abstract:**

Wide proliferation of low temperature hydrogen fuel cell systems, a key part of the hydrogen economy, is hindered by degradation of the platinum cathode catalyst. Here, we provide a device level assessment of the molecular scale catalyst degradation phenomena, using advanced *operando* X-ray scattering tomography tailored for device-scale imaging. Each cell component, including the catalyst, carbon support, polymer electrolyte, and liquid water can be simultaneously mapped, allowing for deep correlative analysis. Chemical and thermal gradients formed inside the operating fuel cell produce highly heterogeneous degradation of the catalyst nanostructure, which can be linked to the macroscale design of the flow field and water distribution in the cell materials. Striking differences in catalyst degradation are observed between operating fuel cell devices and the liquid cell routinely used for catalyst stability studies, highlighting the rarely studied but crucial impact of the complex operating environment on the catalyst degradation phenomena. This degradation knowledge gap highlights the necessity of multimodal *in situ* characterization of real devices when assessing the performance and durability of electrocatalysts.

**One Sentence Summary**: *In situ* correlative X-ray diffraction imaging allows assessment of cathode catalyst deterioration inside operating low temperature hydrogen fuel cells, showing the influence of macroscopic cell design and the operational environment on the nanoscale degradation of the electrocatalyst.

## Main Text:

Large scale adoption of hydrogen energy systems requires reliable, cost competitive water electrolyzers and fuel cells, of which the polymer electrolyte membrane (PEM) variants are most popular (*1*). The performance and cost are both driven by design of a multilayered nanocomposite membrane electrode assembly (MEA) (*2*). Despite great progress in optimizing individual components of the PEM fuel cell (PEMFC) MEAs in laboratory model systems(*3*, *4*), these advances have not always translated into improved performance at the device level (*5*), because strong interactions between different components and the complex chemical environment of an operating cell hinders the optimization of MEA architecture (*6*, *7*). Improving the durability of high performance catalysts presents a great challenge and is limited by our understanding of the degradation phenomena (*8*, *9*). Carefully balancing the performance, durability, and cost of the oxygen reduction reaction (ORR) catalyst is necessary for technological advancement of PEMFC systems (*10*).

The degradation processes of Pt-based ORR catalysts have been rigorously investigated under idealized laboratory conditions (i.e. in inert atmosphere, room temperature, and ultrahigh purity liquid electrolyte). The oxidation and reduction of the Pt surface plays a central role (*11*) with potential cycles triggering nanoparticle reconstruction and dissolution (*12*). This leads to corrosion, ripening and coalescence of the nanoparticles during operation, lowering the surface area and mass activity of the catalyst (*13*). The extent to which catalyst activity and durability measurements under idealized conditions transfer directly to the hot, gas phase polymer electrolyte MEA environment, remains an open question (*14*), and the answer will guide development of active materials and operating protocols towards more robust performance.

Techniques which can simultaneously probe all the components of functional devices at the nanoscale are urgently required to optimize the MEA architecture and operating conditions. Spectromicroscopy approaches using electron (*15*), neutron (*16*), and photon beams (*12*, *17*, *18*) are powerful, but generally laborious and/or limited to *ex situ* experiments. Significant advances in spatial and time resolution with *in situ* X-ray absorption tomography have been recently achieved (*19*, *20*). However, spatial resolution is often limited to scales greater than 100 nm and lacks the critical nanoscale and chemical information about the catalyst and supporting materials. This information is necessary to connect the catalyst activity, morphology and stability to the overall MEA performance (*21*). These properties determine the device's life cycle, with major impact on its economic feasibility (*22*).

Here, we study the catalyst degradation phenomena in an operating PEMFC using novel high energy X-ray scattering techniques, simultaneously mapping chemical composition and nanostructure across the device scale of several centimeters (*23*). Major advancements in computed tomography (CT) algorithms allow a direct observation of the rapid heterogeneous nanostructural deterioration of the MEA. The catalyst degradation is directly linked to the formation of chemical and temperature gradients, correlated to the water distribution in the cell materials, and imprinted by the macroscale geometry of the flow field. Characterizing and predicting the gradients at different operating conditions and integrating the structural evolution of electrocatalysts with fuel cell design will be increasingly necessary to mitigate degradation and lower the cost of the technology.

To simultaneously probe the operating cell at multiple length scales, a combination of synchrotron X-ray diffraction computed tomography (XRD-CT) and small angle X-ray scattering computed tomography (SAXS-CT) was employed (Fig. 1A) (*24, 25*). XRD-CT has been previously used to investigate solid oxide fuel cells (*26*), catalytic reactors (*27*) and batteries (*28*) from hundreds of micrometers up to millimeters in size. To understand the degradation phenomena at the device level, we have extended these techniques to samples 10x larger than previously possible, such that entire X-ray transparent 5 cm$^2$ PEMFCs (*29*) can be imaged (Fig. 1B). The voxel size (200 µm x 200 µm x 50 µm) is currently limited by the beam size and measurement speed, and can be optimized for larger devices down to the sub-micrometer level using 4th generation X-ray sources. Detailed descriptions of the method are available as supplementary information.

The main advantage of XRD-CT is its superior chemical contrast in comparison to standard absorption-contrast tomography (Fig. 1C). The chemical maps, which can be merged into 3D images (Fig. S1), reflect the intrinsic heterogeneous nature of the PEMFC largely imprinted by the flow field design (Fig. 1B). Compression of the flow field induces wrinkling in the MEA, on the order of a few microns, generating lined patterns in the Pt and ionomer signals (note that the parallel channels of the cathode and anode plates were aligned perpendicular to each other)(*30*). This heterogeneous pressure field alters mass and thermal transport, as well as the mechanical properties of the electrodes, generating chemical gradients inside the cell and defining the nanoscopic degradation phenomena in the catalyst layer, which we will show later.

One of the most important aspects of fuel cell operation affected by the flow field is the water transport and distribution inside the catalyst and porous diffusion layers, which is critical for obtaining maximum cell performance (*19, 31*). XRD-CT allows the spatial distribution of liquid water to be obtained from the amorphous background of the XRD patterns (Fig. S5), which has the considerable advantage that the water distribution can be simultaneously correlated with the nanoscopic structure of the ionomer membrane, and carbon support at identical locations during the operation of the device. The water distribution obtained in this manner is shown for slices collected at two different conditions: i) at 1 V, corresponding to an idle cell when no power is drawn, and, ii) at 0.6 V, where the cell is operating at high current density (1 A/cm$^2$, 0.6 W/cm$^2$). Differences in the quantity and distribution of water are detected throughout the device (Fig. 1C, far right) and local water accumulation is observed, especially inside the ionomer membrane under the landing areas where the flow field structure touches the MEA. Unfortunately, the water content in pixels containing large quantities of Pt (i.e. inside the catalyst layer specifically) is difficult to obtain as the scattering from water is masked by the high attenuation of Pt.

The general phenomena of water collecting underneath the landing regions of fuel cell flow fields has been studied for decades using X-ray (*32*) and neutron probes (*33, 34*). Detailed quantitative measurements of ionomer microstructure by SAXS add significant value to standard water imaging experiments, since hydration can be more precisely determined (*35, 36*) and data can be readily compared to the wider body of ionomer literature. Furthermore, SAXS-CT allows the spatial correlation of ionomer micro/nano structure, with the water content in the other cell constituents while using the same X-ray probe. A SAXS-CT slice collected in a separate experiment during cell operation (0.8 A/cm$^2$, 0.59 V, 80% relative humidity) reveals the heterogeneous hydration of the membrane throughout the MEA with significantly better detail then in the XRD measurement (Fig. 2). The SAXS curve in each voxel was modelled as the sum of two components: a power law Porod decay related to macroscale polymer organization and a Gaussian function for the ionomer peak (*37*). Because the thin membrane was not perfectly aligned with the plane of the slice, we restrict our analysis to the regions where the ionomer peak is clearly visible. Comparatively wet and dry regions inside the membrane can be directly resolved through changes in the *d*-spacing of the ionomer peak (Fig. 2A). This indicates the distance between ionic nanodomains increases from 3.5 to 3.8 nm in the preferentially hydrated regions (Fig. 2B). The wet and dry regions exhibit a vertically lined pattern correlated with the positions of the flow field. This pattern is similar to the water distribution from the XRD-CT measurements, validating the data obtained by less precise XRD pattern analysis (note that the cathode and anode flow field ribs were aligned parallel to one another for the SAXS imaging). The peak position shows that the membrane is far from completely hydrated, about 10% by weight (fully hydrated membranes contain up to 25% water (*38*, *39*)). Streaks in the SAXS-CT image show the membrane underneath the flow field rib is preferentially hydrated, and far from equilibrium with the neighboring regions inside the cell, pointing to the same water distribution pattern within the membrane as in the carbon diffusion layers (*34*).

To correlate the water distribution within the cell to the nanostructural heterogeneity of the catalyst, linked to the aging mechanisms within the MEA (*40*, *41*), we extract the catalyst's structural parameters from the XRD-CT patterns. The intensity parameter of the Pt diffraction signals corresponds to the quantity of material, while the peak widths are related to the average particle size and defect content of the Pt catalyst. The main complication limiting such analysis on full scale operational devices are peak broadening artefacts known as parallax errors (i.e. peak broadening as a function of tan2θ, Figs. S9 & 10). These distortions in XRD-CT images of large samples >1 cm have until now limited the quality and reliability of nanostructural parameters obtained from the diffraction data (e.g. lattice parameter, peak widths) (*29*). Here we disclose a new tomographic algorithm called Direct Least-Squares Reconstruction (DLSR) which provides a generalized solution to this problem, and corrects for parallax distortion in arbitrarily large samples. Therefore, the practical maximum sample size for XRD-CT is limited only by penetration of the X-ray beam. This advance extends the capabilities of scattering-contrast tomography from laboratory specimens to samples normally imaged using hospital-style CT scanners, including practical fuel cell devices. The DLSR algorithm quantitatively recovers the diffraction profile (Figs. S11-S15) and the microstructure of the Pt catalyst in the fuel cell, including the particle size and strain. The computationally expensive nature of the parallax-corrected reconstruction currently limits the image resolution, although rapid advances in software capability are expected. A detailed discussion on parallax artefacts, its correction using the DLSR algorithm, and benchmarking versus conventional algorithms can be found in the Supplementary Information.

Fig. 3 shows images of the Pt nanostructure obtained from Rietveld analysis of a single XRD-CT image slice collected through the cathode of the MEA during operation at 1 A/cm². The degradation was induced by holding the potential at 1 V for 10 hours at 80°C (*13*), which is slightly above open circuit voltage (ca. 0.95 V). These high, cell reversing potentials lead to Pt oxidation and degradation (*42*). Variations in the Pt intensity map (Fig. 3A), reflect the mechanical distortion of the catalyst coated membrane inside the pressurized cell during operation. The region where the ionomer membrane protrudes into the measured slice shows particles up to 100 nm in size (yellow arrow, in Fig. 3B). While degradation gradients through the cross-section of catalyst layers have been previously reported using *ex situ* TEM (*43*, *44*), the size of the nanoparticles overlapping these protrusions are an order of magnitude larger than expected. Nanoparticles of this size are typically found in the "Pt band", formed near and inside the membrane of aged MEAs (*45*, *46*). Even though the formation of this Pt band has been identified as a key factor in the stability of field-tested MEAs (*47*), *in situ* detection of Pt band formation has not been previously reported.

The central region of the sample (Fig. 3B, magenta arrow) can be seen in greater detail in Figs. 3C & D. Although the spatial resolution of these parallax corrected images is limited (500 µm) the particle size in the sample at the beginning of testing (Fig. 3C) is much more spatially homogeneous than after testing (Fig. 3D). Higher spatial resolution XRD-CT images can be obtained if parallax correction is neglected (Figs. 3E and F, 200 µm resolution), with the tradeoff that the calculated particle sizes are qualitative and only comparative analysis is possible. Vertical and horizontal streaks are visible in the uncorrected maps, where the particles are systematically larger in regions covered by the flow field plates. Therefore, the morphological degradation of catalyst nanoparticles in these regions is more severe, coincident with previously observed thinning of the membrane and local Pt corrosion (*48*, *49*). The specific factors responsible for this catalyst nanoscale heterogeneity under near-zero load are not easily decoupled, but our data suggest correlation to the water distribution in the cell, temperature, and mass transport of dissolved species. A detailed understanding of how these chemical gradients precisely control ageing processes remains unresolved, even though the degradation mechanisms of Pt catalysts have been investigated extensively in both nanoparticle model systems and full MEAs (*12*, *21*, *45*, *46*, *50*).

Given the heterogeneity observed inside the catalyst layer during ageing in an operating PEMFC device, the relevance of ageing data from laboratory model systems deserves scrutiny because catalyst durability is principally assessed using cycling in ultrahigh purity liquid electrolyte at room temperature on a rotating disc electrode (RDE) (*51*, *52*) . To investigate the possible differences, the most popular catalyst accelerated stress testing (AST) protocol (sweeping between 0.6-1.0 V at 50 mV/s with Ar atmosphere on the cathode) was employed on the same commercial catalyst (Fig. S21) inside the PEMFC and with a conventional three electrode RDE cell (*2*). X-ray diffraction was measured *in situ* at multiple locations in the cathode throughout the AST of the PEMFC sample and transmission electron microscopy (TEM) images were collected for both the PEMFC and the RDE cell catalyst at the beginning and end of the test.

Both catalyst particle ripening and dissolution of Pt in the PEMFC sample are directly observed during the AST. The mean crystallite size of the Pt increases from 4.6 nm to 10.3 nm (Fig. 4, top) and approximately 30% of the Pt in the cathode dissolved (Fig. 4, bottom), which is consistent with previous work (*12*, *42*). Neither parameter follows a linear trend with the number of cycles: a large fraction of the overall change takes place in the first few cycles and the degradation slowly diminishes in effect without reaching a plateau (*40*).

As anticipated, the catalyst in the PEMFC aged dramatically faster than inside a conventional RDE cell. These differences can be attributed to several factors, including temperature, pH, and mass transport of dissolved species. Although no tomography was performed, XRF mapping of the MEA sample after ageing showed that the dissolution was fairly homogeneous and XRD at multiple locations, both in the channels and landing areas, yielded similar results (Fig. S26). These data indicate that the chemical environment created inside the fuel cell during this AST profile is more homogeneous than even an idle operational cell, and questions the utility of current ASTs to assess catalyst stability during PEMFC operation. This is because the chemical and temperature gradients, which define the degradation dynamics as discussed above, are linked to faradaic reactions, and are suppressed during stress testing under inert atmosphere. Therefore, the only reliable way to benchmark the catalyst stability, is to study the materials in a PEMFC device during operation. Furthermore, the predictions obtained from laboratory model systems are substantially and systematically different than catalyst degradation in devices.

In conclusion, major advances in XRD-CT reconstruction algorithms allow for simultaneous assessment of nanostructural and chemical heterogeneities inside an operational 5 $cm^2$ MEA for the first time. Interactions between different phases can be correlated at high current densities and with one probe, which is a significant advantage compared to more selective electron or neutron imaging. The XRD tomographs show spatially resolved differences in the ageing of catalyst particles, which correlate to the flow field geometry and water distribution within the device. Therefore, the macroscopic design of the cell has a decisive role in the nanoscopic degradation phenomena of the catalyst, which needs to be taken into account at all levels of PEMFC design. Accelerated stress testing tracked by *in situ* X-ray diffraction indicates that catalyst ageing inside PEMFCs is extremely aggressive when compared to conventional RDE testing, even in the absence of faradaic currents which further accelerate the ageing. Stress tests performed in a liquid cell should be critically evaluated in order to predict catalyst degradation rates and mechanisms in functional systems.

The presented strategy for deep, *in situ* and *operando* characterization bridging fundamental chemistry and device engineering is generally applicable to the next generation of batteries, solar cells, and other energy conversion and storage systems with few compromises in electrochemical performance. We anticipate that such non-intrusive and holistic approaches, which combine several X-ray scattering tomography techniques, will enhance understanding of the roles, and interactions between different constituents, allowing next-generation materials to be incorporated into PEMFCs and other electrochemical devices.

.

**Acknowledgments: Funding:** This work was supported by Automotive Partnership Canada and NSERC. A.V. is supported through the European Union Horizon 2020 program under Grant Agreement No. 679933 (MEMERE). The SAXS-CT was funded by Institut Carnot Energies du Futur under the STATUQUEAU project. **Author contributions:** I.M., A.V., R.C., J.P., M.V.B., E.A.F., T.A., T.S., D.B., D.P.W, and J.D. performed the XRD experiments. N.M, S.E., F.M., S.L., A.M, and J.D. performed the SAXS experiments. A.V., A.C., and S.J. devised and implemented the DLSR algorithm. I.M. and J.D. wrote the manuscript. D.B., D.P.W., S.L., A.M., V.H., S.J., F.M., L.D. and J.D. supervised the work. All authors reviewed the manuscript. **Competing interests:** Authors declare no competing interests. **Data and materials availability:** All data are available upon reasonable request from the corresponding authors.

**Supplementary Materials:**

Materials and Methods

Figures S1-S36

References 52-70

**Supplementary Information**

**Materials and Methods**

**3D reconstruction**

It is possible to assemble 3D chemical maps using XRD-CT data. While only three XRD-CT slices were measured here, a 3D representation of the chemical maps from Fig. 1C in the main article text was assembled, and can be seen in Fig. S1. This analysis allows for easy correlation between features in multiple phases. Thresholding and segmentation are often required to analyze 3D datasets in a visually interpretable fashion. These processes designate each voxel to a specific phase, minimizing overlap. With diffraction contrast tomography, this step is not required, but has been used here to produce a simplified image.

**Membrane electrode assemblies and cell**

A membrane electrode assembly (MEA) purchased from BioLogic was used for the X-ray Diffraction-Computed Tomography (XRD-CT) experiment (Figs. 1 and 3, in the main article text). The catalyst coated membrane was a Nafion®115 film, coated with Pt/C ($0.50 \pm 0.05$ $mg_{Pt}$/$cm^2$) on both cathode and cathode. The gas diffusion media were not specified, but appeared to be a carbon cloth supporting a microporous layer.

A Solvicore H600 MEA was used for the Small Angle X-ray Scattering-Computed Tomography (SAXS-CT) experiment (Fig. 2 in the main article text). The 20 µm thick reinforced ionomer film was coated with Pt catalyst supported on graphitized carbon for the cathode, and Pt supported on carbon black for the anode. The catalyst loading was not specified. The gas diffusion layers were not specified.

A custom MEA from Johnson Matthey Fuel Cells was used for accelerated stress testing of the catalyst (Fig. 4, in main article text). The catalyst on each electrode was not directly specified, but the performance, elemental analysis, XRD, and electron microscopy is consistent with the Johnson Matthey HiSpec 4000 catalyst (40 wt % Pt on Vulcan XC72R). The specification for maximum crystallite size for this catalyst by XRD is 4.5 nm. The catalyst loading on the cathode was specified as 0.44 $mg_{Pt}$/$cm^2$, and the anode loading was 0.04 $mg_{Pt}$/$cm^2$. The ionomer membrane was a proprietary short side chain, fiber reinforced, 17 µm thick perfluorosulfonic acid membrane. Double sided catalyst coated membranes (CCMs) were obtained directly from the manufacturer and assembled into MEAs by the authors. The 3-layer catalyst coated membranes were hot-pressed (60 MPa, 60 s, 120°C) against Sigracet 25BC (Fuel Cells Etc) gas diffusion layers with microporous layers on the cathode and anode to produce a 7-layer MEA. Adhesive polyimide film was used to frame the MEA and provide a gas-tight seal for the Viton gasket.

A schematic of the flow fields used in the cell for the XRD-CT experiment can be seen in Fig. S2. For the SAXS-CT and accelerated stress testing experiments, the same flow fields were aligned in a parallel orientation (rotating the cathode by 90°). More detailed drawings of the cell design have been reported elsewhere (*29*).

**X-Ray Diffraction**

The XRD and SAXS patterns were radially integrated using PyFAI(*53*) $CeO_2$ (NIST, SRM 674b) was used as an area detector calibration standard for the XRD, while Ag behenate was used for the SAXS. A Pilatus 2M CdTe detector was moved between two positions to capture the XRD and SAXS data. An evacuated flight tube was used to reduce parasitic scattering in the SAXS geometry. The Rietveld refinement on accelerated stress test measurements was analyzed using GSAS-II(*54*), as described previously (*55*). Fig. S3 shows the Rietveld refinement of the in situ XRD collected from the fresh catalyst before accelerated stress testing. The lattice parameter, particle size, and scale factor of the diffractogram were refined. The Debye-Waller factor for Pt, and approximate instrumental broadening parameters were calibrated using a polyimide capillary of Pt/C catalyst material, and the $CeO_2$ standard, respectively. The Rietveld refinement was performed over the first 17 reflections of the Pt lattice, from 3.0 to 16.5 degrees $2\theta$. An 8 term Chebyschev polynomial background was modelled. Fitting the diffractograms with additional parameters, such as microstrain, anisotropic particle sizes, or Voight broadening contributions was attempted. No substantial improvement in the fit quality was observed. A schematic of the slightly tilted (0.3°) cell geometry used for the accelerated stress test and its influence on the apparent XRD line shape is visible in Fig. S4. In this configuration the parallax effect can be suppressed.

Fig. S5 shows the typical reconstructed XRD pattern for one voxel near the center of the fuel cell, collected at a cell voltage of 0.6 V. This pattern can be decomposed into a linear combination of all the MEA components: Pt catalyst, PFSA ionomer membrane, carbon support, carbon from the gas diffusion layer, and also water. This is done using reference patterns for each constituent materials and an optimization algorithm. This is a very sensitive method for quantifying the amount and distribution of these different materials, as the XRD patterns are unique for each constituent. The background is characterized by a constant function in the fitted region as most of the background signal coming from the cell parts is subtracted during the reconstruction process. The negative going features seen at low angle are reconstruction artefacts which imperfectly subtract the strong background scattering signal from the PEEK sample environment casing. These reconstructed diffractograms are suitable for extracting peak intensities, but not for high-quality peak shape analysis, because of the parallax errors discussed below.

Fig. S6 shows a replotting of the accelerated stress test data shown in Fig. 4, but with logarithmic scaling for the number of potential cycles.

Fig. S7 shows the spatial heterogeneity present inside the cell during accelerated stress testing under inert conditions. Diffraction from multiple positions was measured after each number of stress test cycles. The data in Fig. 4 correspond to the catalyst underneath the length of a cathode flow field rib in the center of the cell (Fig. S7, black squares, position 1). The catalyst down the length of the adjacent channel was also measured (red squares, position 2). The most significant difference between these two datasets is the calculated lattice parameter (Fig. S7A). The lattice parameter is very stable for the catalyst under the flow field rib (within $10^{-3}$), while the apparent lattice parameter for the channel region exhibits large, erratic fluctuations over time. These fluctuations are an artefact from mechanical distortion and parallax error. The catalyst layer pressed against the flow field maintains its mechanical rigidity, while the portion of the MEA floating freely in the middle of the channel twists, bends, and distorts over the 48 hours of the cycling. As different portions of the catalyst layer move in and out of the beam, the average detector sample distance of the illuminated volume is altered, which produces an apparent shift in the lattice parameter. This movement also couples to the diffracted intensity (Fig. S7B), which is notably less stable for the catalyst in the channel region. These shifts in peak position and intensity do not strongly couple to the peak widths. Therefore, the particle size of the catalyst can be measured for both positions, and the same trend is observed as a function of stress testing cycles (Fig. S7C).

Wrinkling and bowing of the MEA can be visualized by scanning the beam vertically through the cross section of the catalyst layer. For a perfectly planar MEA, the XRD peak profile should remain constant, only changing in intensity. In practice, the profile depends on the vertical beam position (Fig. S8). Peak broadening is visible for measurement positions above and below the central, aligned position of the catalyst layer, indicated by the black line. The flow field ribs do not run the entire length of the sample. In the inlet and outlet regions, the membrane is not held flat by the flow field plate, and is distorted. In these locations, the parallax effect is strong enough to split the reflection into two separate peaks at 6.45° and 6.7°. This effect is corrected in the tomography analysis. Tightly focused X-ray beams and precise alignment are necessary for measuring high quality diffraction on these samples.

**Parallax Distortion in XRD-CT**

In X-ray scattering/diffraction computed tomography experiments, it is assumed that the X-rays scattered/diffracted from the sample at a certain $2\theta$ angle arrive at the same detector element/pixel. This is illustrated in Fig. S9.

Typical transmission XRD experiments assume that the detector is in the "far-field", where the area detector and detector sample distance are much larger than the sample. This optical approximation is useful where the sample thickness is on the order of 1 mm, and the detector sample distance is at least about 0.5 m. However, when the sample diameter increases to tens of mm, then this assumption is no longer valid. For large samples, X-rays diffracting at the same $2\theta$ angle will arrive at different detector elements, as the sample-to-detector distance varies significantly from the back to the front of the sample. This phenomenon is known as the parallax effect and has a $\tan(2\theta)$ dependence (56). The parallax effect is illustrated in Fig. S10.

The artefact takes form as peak shifting and broadening, but in many cases can even lead to peak splitting. Recently Stock *et al.* suggested a simple approach to overcome the parallax artefact's influence on measured lattice parameters (57):

1. A translation scan is performed at an angle $\varphi$

2. The same translation scan is performed at an angle $180° + \varphi$

3. Full profile analysis of the acquired diffraction patterns in order to extract the lattice parameter values for the structural model

4. The mean value of the lattice parameter values for each position is calculated using the diffraction patterns from the two angles

Unfortunately, although not stated by the authors, this approach is only applicable to homogeneous samples where a single unit cell can be used to model the full scattering data. In the case illustrated in Fig. S10, the sample has two different components: a dark blue region with small crystallites, and a light blue region containing larger crystallites of the same phase. It becomes necessary to use two copies of the same unit cell to separately model the diffraction patterns of the dark blue and light blue components. It can be easily understood that the number of structures to be used will vary at different positions, and scales arbitrarily. For the same reasons, the information regarding the peak width (e.g. crystallite size) is also lost. As we have previously noted, this is also why direct peak fitting/modelling of the projection data of heterogeneous samples using one model per pattern can lead to wrong results, and indeed wrong conclusions. For example, if there is a region in the sample along the beam path where a specific phase is generating broad peaks, and in a different region generating very sharp diffraction peaks, a single average physical model is not sufficient to describe the peak shape in the recorded projected diffraction pattern.(58)

The phenomenon described here should not be confused with a different effect sometimes referred to as parallax error, associated with X-ray photons traversing through multiple detector pixels at high $2\theta$ angles, in detectors with thick sensors. This effect, also known as detector penetration, results in shifts and broadening of the diffraction peaks, but it has been recently demonstrated by Marlton et al. that the application of a simple quadratic polynomial offset can model the observed $2\theta$ shifts and correct for these errors (59).

**Direct Least-Squares Reconstruction**

We have developed a novel XRD-CT data reconstruction algorithm that we term the Direct Least-Squares Reconstruction (DLSR) algorithm, which to date provides the only comprehensive solution to the parallax artefact. It should be noted that the DLSR algorithm is generic, and can be applied to any chemical tomographic technique. The only requirement is an available mathematical model that accurately describes the artefact-free observed spectroscopic/scattering data.

After the initial reshaping (and radial integration in the case of 2D diffraction/scattering data), the raw observed data in a chemical tomography dataset constitute a 3D matrix. It is a stack of 2D images (sinograms), each one corresponding to one spectral/scattering channel. Conventionally, the data analysis approach for such chemical tomography datasets is the "reverse analysis" algorithm(60):

1. A data volume is reconstructed using each of the sinograms in the stack. Each voxel in the reconstructed data corresponds to a local spectrum or diffraction/scattering pattern.

2. Perform single/multiple peak fitting or full profile analysis for each voxel independently, yielding local physico-chemical information.

The DLSR algorithm merges these two steps into one. To clarify, a chemical tomography dataset consists of nT translation steps, nP tomographic angles and nCh channels corresponding to the spectral/scattering dimension; it is a 3D matrix with dimensions nT x nP x nCh. In the "reverse analysis" approach, the reconstructed volume is also a 3D matrix with nT x nT x nCh dimensions, an 2D image slice where each voxel contains a diffractogram/spectra. The resulting maps containing the extracted physico-chemical information can be modeled as a 3D matrix with nT x nT x nPar dimensions, where nPar are the number of refined parameters in each voxel (e.g. peak(s) position/width/shape, scale factors, lattice parameters and crystallite sizes). In contrast, the DLSR algorithm uses the raw sinogram volume (3D matrix with (nT, nP, nCh) dimensions) and yields directly the images containing the local physico-chemical information (3D matrix with (nT, nT, nPar) dimensions).

Iterative algebraic reconstruction algorithms in computed tomography yield reconstructed images from sinograms by solving a system of linear equations A x = b, where (61, 62):

- A is the coefficient matrix corresponding to ray tracing during the tomographic scan (conventionally a 2D matrix with (nT x nP, nT x nT) dimensions)

- x is the solution matrix (reconstructed image, conventionally as flattened 1D matrix with (nT x nT, 1) dimensions)

- b is the matrix which corresponds to the observed data (sinogram, conventionally as flattened 1D matrix with (nT x nP, 1) dimensions)

Inspired by this, the DLSR algorithm solves an A S x = b system of non-linear equations using a non-linear least-squares minimization approach, where:

- A is again the coefficient matrix corresponding to ray tracing during the tomographic scan and a 2D matrix but with ( nT x nP x nCh, nT x nT x nCh) dimensions.

- S is a 2D matrix with (nT x nT x nCh, nT x nT x nPar) dimensions

- x is the solution matrix corresponding to images containing the physico-chemical information and in a flattened 1 D matrix form with (nT x nT x nPar, 1) dimensions

- b is the matrix which corresponds to the stack of sinograms as flattened 1D matrix with (nT x nP x nCh, 1) dimensions)

In practice this means that nT x nT physical models are created to describe the nT x nP x nCh observation points. This is a computationally intensive approach as it is many non-linear equations with shared parameters to be solved simultaneously.

**Simulated XRD-CT Data Without Parallax Correction**

The DLSR method was first benchmarked against the conventional reconstruction technique using simulated data. In this work, we implemented the DLSR algorithm for XRD-CT using the Topas Academic v7 (64 bit) (*63*). First, we simulated a phantom XRD-CT object using a Pt unit cell (nT = 81, nP = 81, nCh = 1680) and performed data analysis using both approaches ( i.e., the reverse analysis method and the DLSR algorithm). The simulated data contained no noise or parallax distortion, in order to compare the performance of the two approaches under ideal circumstances:

1. Three Pt diffraction patterns were simulated (nCh number of scattering points) by varying the scale factor, lattice parameter and crystallite size.

2. A phantom (nT, nT) image was created, and three regions of interest were segmented.

3. A volume with (nT, nT, nCh) dimensions was created, and assigned one diffraction pattern at each region.

4. The volume was forward projected (using the A coefficient matrix from the Airtools package(*64*)), to create the artificial XRD-CT sinogram data.

5. For the conventional approach, the XRD-CT images were reconstructed using the filtered back-projection (FBP) algorithm and then each diffraction pattern was analyzed independently using the Topas software. For the DLSR algorithm, a Topas.INP file was created taking into account the coefficient values from the A matrix and creating in total nT x nT Pt unit cells; simultaneous Rietveld analysis of nT x nP diffraction patterns with nCh bins using nT x nT Pt unit cells with shared parameters led to nPar images with (nT, nT) size. In this case, nPar = 3, corresponding to the Pt scale factor, lattice parameter and crystallite size.

The results are presented in Fig. S11. The ground truth (GT) corresponds to the simulated dataset. Both the DLSR and conventional reverse analysis FBP algorithm, produce good fits to the data. The results from the DLSR algorithm are slightly superior to the ones obtained with reverse analysis. This can be observed from the Pt scale factor maps but more characteristically from the Pt lattice parameter and crystallite size maps. In the latter two, the reverse analysis method seems to be prone to some minor reconstruction artefacts (maximum error of 0.01 Å in lattice parameter and several nm in crystallite size in certain regions) while the DLSR algorithms yields excellent results (overall $R_{wp}$ = 0.35 % with these noiseless patterns). Under ideal conditions, both methods accurately reconstruct the simulated data. The DLSR is marginally better in reconstructing all parameters, and this new methodology does not require any compromises in the data quality.

**Simulated XRD-CT with Parallax Correction**

Next, the two methods were compared using a simulated dataset containing parallax distortion. The basic concept behind our algorithm that overcomes the parallax artefact lies on the idea that the sample can be represented by a grid where each cell corresponds to a unique sample-to-detector distance. The grid contains nT x nT cells, equal to the size of the reconstructed images. The sample-to-detector distance changes from the back to the front of the sample with a constant step which is equal to the translation step size during the tomographic scan. This is illustrated in Fig. S12, where it is also demonstrated that the Q axis calibration at the front of the sample is different from the scattering from the back of the sample.

The XRD-CT scans of the polymer electrolyte membrane fuel cell (PEMFC) consisted of 230 translation steps (translation step size of 200 µm) covering a total length of 4.6 cm during each translation scan. The detector calibration was performed using a single point measurement of $CeO_2$ calibrant (NIST, SRM 674b) at ca. 500 mm distance from the detector. The diffraction data were integrated with pyFAI using a point of normal incidence calibration file. In total, 230 scattering vectors ($2\theta$ axes) were calculated with pyFAI by altering the sample-to-detector distance accordingly (steps of 200 µm). This led to the creation of the Scattering Displacement Grid (SDG) which is a 3D matrix with (nT, nT, nCh) dimensions. The SDG was downsized to nT = 81, due to memory restrictions. To simulate the artificial XRD-CT that contains parallax artefacts, the following strategy was implemented:

- For every tomographic angle, the SDG was rotated by the same angular value

- The original Pt XRD-CT dataset (Ground Truth in Fig. S11) was interpolated in 3D using the new SDG yielding a new distorted 3D matrix (DM) with (nT, nT, nCh) dimensions.

- The projection data for this tomographic angle were calculated by multiplying the coefficient matrix A with the DM. A new coefficient matrix A was calculated for every tomographic angle using the Airtools package.

- The process was repeated for all nP = 81 angles, yielding an XRD-CT dataset containing parallax artefacts.

The implementation of the DLSR algorithm for data containing parallax in Topas was performed by applying a different $2\theta$ offset for every structure (nT x nT structures in total) for every tomographic angle. The results from the reverse analysis method and the Topas implementation of the DLSR algorithm are presented in Fig. S13. It can be clearly seen that the information regarding the Pt lattice parameter is lost in the reverse analysis method while the results regarding the crystallite size are altered dramatically leading to wrong conclusions. For example, the round central region in the sample that corresponds to 15 nm crystallite size is calculated as less than 5 nm with the conventional reconstruction. In contrast, the DLSR algorithm exhibits astonishing robustness, yielding images with correct physico-chemical information. The downside of the DLSR algorithm is the huge requirements in memory (e.g. >20 GB RAM for nT=81, nP=81, single phase, and noiseless patterns). Because the diffractograms from each voxel are not independently reconstructed, it is not possible to manually inspect the fit quality in the same fashion as standard Rietveld refinement. Therefore, the parameter fitting needs to be performed with care, ideally with pre-existing knowledge of the sample.

**Parallax Correction of Experimental PEMFC Dataset**

As a final step, the Topas implementation of the DLSR algorithm was tested using the experimental PEMFC XRD-CT data. The results from an XRD-CT dataset under open circuit voltage (OCV) and the corresponding ones obtained with the reverse analysis method are presented in Fig. S14. Impressively, the results follow the same pattern as the simulations. It can be seen that the Pt crystallite size is severely underestimated with the reverse analysis method. This is due to the peak broadening from the parallax effect. The reverse analysis method yields Pt lattice parameter maps showing gradients where they should not exist, while the DLSR algorithm leads to more uniform and indeed realistic Pt lattice parameter values.

The refinements were performed using the development version 7 of Topas Academic using the FP peak shape function and the Approximate_A function to handle the large scale refinements. The errors in each parameter were calculated by performing 50 cycles of bootstrap_errors. Fig. S15 shows the refined values for the scale factor, lattice parameter, and crystallite size before and after aging, together with maps of the estimated error. These error maps allow the accuracy of the XRD-CT fitting to be quantitatively assessed.

**Small Angle X-ray Scattering**

The SAXS-CT measurement was performed using the same fuel cell sample environment, energy, and beamline configuration as the XRD-CT. The fuel cell was operated at 0.8 A/cm$^2$ during the data acquisition, which corresponded to 0.585 V. Both sides of the cell and gas streams were held at 80% RH and 75°C. The gas pressure on both sides was 150 kPa. The flow rates of H$_2$ and air were 1200 and 3000 standard cubic centimeters per minute (sccm). These conditions reduce the quantity of liquid water condensed outside the PFSA membrane, and lead to stable performance. The inlets of the cathode and anode were configured opposite to one another to produce counter current flow.

Because of limitations in the detector dynamic range, the beam was attenuated 10-fold, and the exposure time of each point was 3 s. The data was reconstructed using the Simultaneous Algebraic Reconstruction Technique algorithm in order to bound the reconstruction to positive values (*65*). A mask was then applied to the reconstruction in order to isolate the ionomer membrane from the other components (Fig. S16). The SAXS patterns were then fitted with the sum of a power law ( Porod decay) accounting for the interfacial regions of very large objects, and a gaussian function representing the nanostructural segregation of the ionic domains. The position of the ionomer peak was then used to compute the d-spacing via the classical relation d=$2\pi/Q_0$. This parameter allows us to quantify the hydration state of the membrane: higher d-spacing values indicate a higher degree of nanoscopic swelling, e.g. higher hydration and larger hydrophilic domains. An example of the hydration levels on the SAXS patterns are presented in Fig. S16B. This data is in good agreement with previous work indicating fuel cell operation causes membrane regions under the current collecting ribs to exhibit higher levels of hydration (*66, 67*). This also aids in the discrimination of the channels and landed areas, shown in Fig. S16C.

While in principle, SAXS-CT also suffers from parallax distortion, this effect is insignificant for two reasons. First, the low angles in SAXS produce a smaller projected distortion. Second, the distortion scales with the ratio of the sample thickness to detector distance, which is at least ten times larger in SAXS geometry than in conventional XRD (6.5 m versus 0.5 m, in this case). Therefore, no parallax correction was necessary for the SAXS-CT images.

**Transmission Electron Microscopy**

A JEOL 2010 transmission electron microscope (TEM) was used to measure the particle size distribution (PSD) of the carbon-supported Pt nanoparticle catalysts used in each experiment (Figs. S17-S23). Catalyst particles were prepared by depositing a drop of isopropanol on the electrode, then transferring and drop casting this solution onto a holey-carbon TEM grid. Commercial Pt/C catalysts are well-known to exhibit significant polydispersity. Several images were collected at different magnifications to understand the heterogeneity of each sample. It was noticed that the PSDs were not perfectly homogeneous when imaging different locations on the TEM grid. Therefore, the particle sizes from images collected at three different locations were averaged when calculating the PSD of each catalyst. The three high resolution images used for particle size measurement of each catalyst are shown, as well as lower magnification survey frames. All of the identifiable catalyst particles in the high resolution images were counted, to preclude any bias in particle selection. Several hundred particles were counted in each case to provide sufficient statistics, and are noted in the individual figure captions. For aggregated particles where multiple crystallites could be clearly discerned, the size of each crystallite was counted individually. Image analysis was performed in ImageJ (*68*), and the histograms were fitted using OriginPro software.

The PSD of each sample was modelled with a log-normal distribution, which provided a good fit except for the catalyst from the MEA used for the XRD-CT measurement. (Figs. S20, S23). The center of gravity for each log-normal fit was reported as the particle size in Fig. 4. The difference between number-weighted and volume weighted histograms for these small particles can be seen in Fig. S22A and B.

For the XRD-CT sample, both the polydispersity, and the fraction of Pt particles in large aggregates are significantly higher (Fig. S23). The local polydispersity can be represented in the frequency distribution plot, where local populations containing both very small and very large particles are observed. Estimating the size of each crystalline grain inside large aggregates of Pt nanoparticles is extremely difficult using conventional TEM.

**X-Ray Fluorescence**

X-Ray Fluorescence (XRF) maps were acquired *ex situ* using a Fisher XDV-SDD microscope equipped with a polycapillary optic and a eucentric silicon drift detector (Fig. S184). The collimating aperture of the incident beam was 200 µm in diameter, which creates a spot size of approximately 220 µm on the sample. The tungsten tube source was operated at 50 kV and 1 mA with no filtration of the incident or detected beams. 15 s exposures were collected for each point spectra, 200 µm pixel spacing. No intensity or scattering correction was performed. The energy calibration of the detector was verified using high purity Pt foil prior to the measurement. The illumination and fluorescence collection was recorded passing through the gas diffusion and microporous layers of the cathode. The MEA was taped flat to an inverted high density polyethylene weigh boat to eliminate background signal from the sample substrate.

A representative single point, *ex situ* XRF spectra from the center of an MEA is shown below. (Fig. S195) Red lines indicate the expected intensities and positions for a metallic Pt reference. The internal attenuation of fluorescent X-rays travelling back through the carbon gas diffusion layer biases the relative intensity of each peak. Nevertheless, several peaks corresponding to the Pt L edge (8 keV to 14 keV) are prominent and can be quantified. The incident high energy X-ray beam can penetrate the whole electrode, including cathode and anode. The fluorescent photons from the anode are additionally filtered by the thickness of the ionomer membrane. We estimate that the attenuation length of ionomer films at the primary Pt peak (9.4 keV) is approximately 600 µm, much longer than the actual film thickness of 125 µm. For the symmetrical anode/cathode catalyst loading MEA sample measured with XRD-CT, the XRF signal therefore reflects a mixed signal from both the anode and cathode Pt. For the MEA used in the accelerated stress test, the low loading on the anode (11x lower) strongly weights the observed XRF signal towards the cathode.

The XRF map of the sample subjected to accelerated stress testing is shown in Fig. S206. The heat map of the image corresponds to the calculated quantity of Pt at each point. The region of the electrode represented in the XRF map is indicated by the red rectangle overlaid on the photo of the MEA (inset). Several long features yielding high intensities of Pt are detected. These result from wrinkling of the catalyst layer and cracks in the microporous layer of the carbon fiber paper (*69*). High pressure hot pressing of the extremely thin CCM (17 µm ionomer film) against the cracked microporous layer creates intrusion of the CCM into the voids of the microporous layer. Fluorescent X-rays generated underneath cracks in the microporous layer are not filtered by that carbon layer, and are detected more efficiently. The Pt signal variation in the regions that are not beneath the cracks in the microporous layer is fairly small, approximately ±15%. No strong features correlated with the cathode or anode flow field ribs are detected. Analysis of the bare, unpressed CCM did not reveal any wrinkling artefacts, and gave homogeneous intensity at all locations.

The XRF map of the sample used for XRD-CT shown in Figs. 1 and 3 was also collected after cell failure (Fig.7). This MEA used a much thicker membrane, and was commercially pressed, likely at lower pressure. No defects in the catalyst film resulting from wrinkles are visible. A horizontal gradient of Pt loading is seen between the anode inlet and outlet, but no features clearly associated with the flow field ribs can be detected.

## Electrochemistry

Before the stress test was performed on the MEA, the cell was briefly conditioned, and a polarization curve was collected. Initial conditioning of PEM cells improves and stabilizes cell performance (*70*). The initial conditioning was performed by gradually increasing the current density of the cell in 200 mA/cm² steps from 0 to 1 A/cm². The cell was then operated at 1 A/cm² for 40 minutes. During this period, the air and hydrogen gas flow were kept at 100% relative humidity (RH), to maximize the hydration of the MEA. These conditions (particularly in 5 cm² cells) produce oscillations in the cell potential over time. These potential oscillations are associated with flooding of the MEA, locally blocking oxygen transport. The cell performance during the initial conditioning phase is visible in Fig. S21. The fluctuations in the applied current density result from a low voltage limit (0.45 V) which improves stability against flooding events in small cells.

A polarization curve was then collected by stepping the current density of the cell in increments from 0 A/cm² to 1 A/cm² (Fig. S9). The cell was kept at 80°C and atmospheric pressure, with 200 and 400 sccm of hydrogen and air flowing through anode and cathode, respectively. The gases were preheated to 80°C and humidified to 100% RH. While these flow rates are both higher and not in the typical stoichiometry of conventionally PEMFC reactant feeds, they produce reasonable stability in this cell and suffice to demonstrate the catalyst performance. At even higher current densities, we have observed the X-ray transparent cell can experience overheating and enhanced degradation.[1] The cell potential was allowed 10 minutes to stabilize at each point. The polarization curve after the stress testing was performed using the same protocol. The results from the polarization testing are shown in Fig. S30. The cell performance before the stress test is equal or significantly better than the performance afterwards, at all current densities.

The electrochemical surface area (ECSA) of the MEA cathode catalyst layer was measured periodically during the stress test by sweeping the cell potential from 0.05 V to 0.60 V at 20 mVs⁻¹ (Fig. S1). The cell current collectors were kept at 80°C, while purging the cathode with nitrogen gas (80°C, 100 % RH, 400 sccm). The anode was purged with hydrogen (80°C, 100 % RH, 200 sccm). These settings were also held constant during the accelerated stress test cycling. The capacitance of the double layer was subtracted by extrapolating a straight line over the region of 0.45 V to 0.60 V. iR drop was measured and corrected by a single point high frequency impedance measurement. The ECSA was calculated using the known pseudocapacitance of 210 μC/cm² Pt for a monolayer of adsorbed hydrogen (Fig. S32222). Note that the calculated ECSA was not corrected for the Pt dissolution (Fig. 4, in the main article text). Specifically, these values assume a constant geometric catalyst loading of 0.44 mg$_{Pt}$/cm². During the accelerated stress test, a gradual decrease in the double layer capacitance and the quantity of Pt oxidized and reduced can also be observed (Fig. S3) while cycling the cell between 0.6 V to 1.0 V.

The electrochemical behavior of the catalyst used in the RDE environment was validated using standard electroanalytical techniques, as described elsewhere (*41*). A BioLogic SP-240 potentiostat was used to record the electrochemical data. The electrochemical surface area was measured by integrating the charge passed during stripping of a CO monolayer during cyclic voltammetry (Figs. S34 and S35). The ORR activity was evaluated by measuring the kinetic current at 0.9 and 0.95 V in oxygen saturated electrolyte (Fig. S6). A detailed discussion on the preparation of RDE catalyst films and measurement of the catalyst activity can be found elsewhere (*5*). Separate electrode samples were prepared for 3500 and 10000 cycles, so that TEM examination of the catalyst could be performed without disturbing the catalyst film. The CO stripping before aging was measured for both samples. The accelerated stress testing of the catalyst was performed at room temperature in freshly prepared, argon saturated 0.1 M $HClO_4$ (96 wt%, Suprapur®Merck) diluted in Milli-Q water, with no electrode rotation. The potential cycling was performed between 0.6 and 1.0 V vs the reversible hydrogen electrode (RHE) at 50 $mVs^{-1}$.

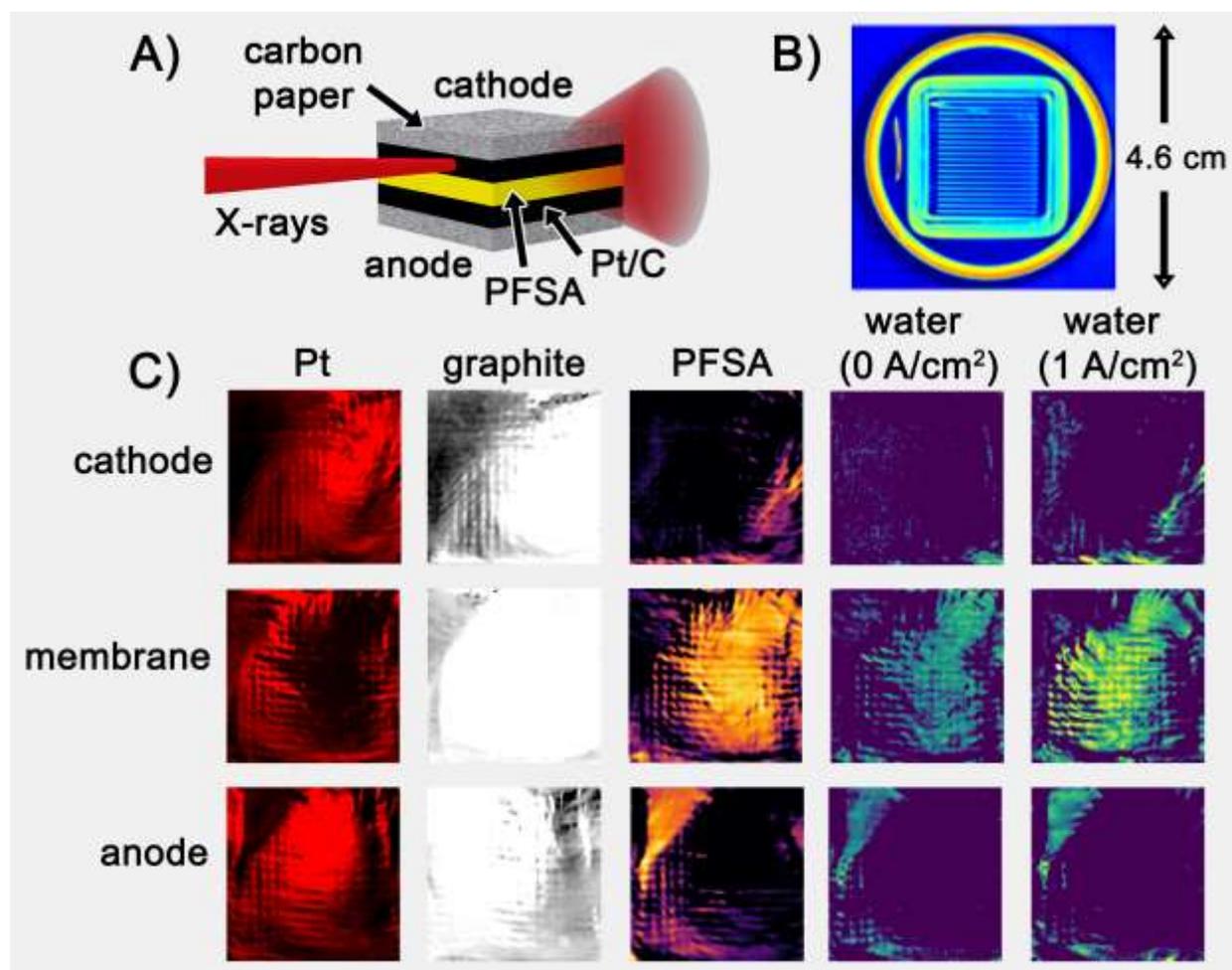

**Fig. 1.** *In situ* XRD-CT of membrane electrode assembly. A) Schematic of tomographic geometry. B) Scattering intensity of slice through circular PEMFC cell casing and square graphite flow field of 5 cm$^2$ cell. C) Chemical maps of tomography slices collected through the cathode, PFSA membrane, and anode and obtained by simple deconvolution of reconstructed XRD patterns (*29*). Each map in C) is 1.9 cm across, 200 µm resolution, measured over the central region covered by the flow field. The cell was operated at 80°C and 100% relative humidity. The three slices were collected at 1 V, and during operation at 0.6 V, 1 A/cm$^2$.

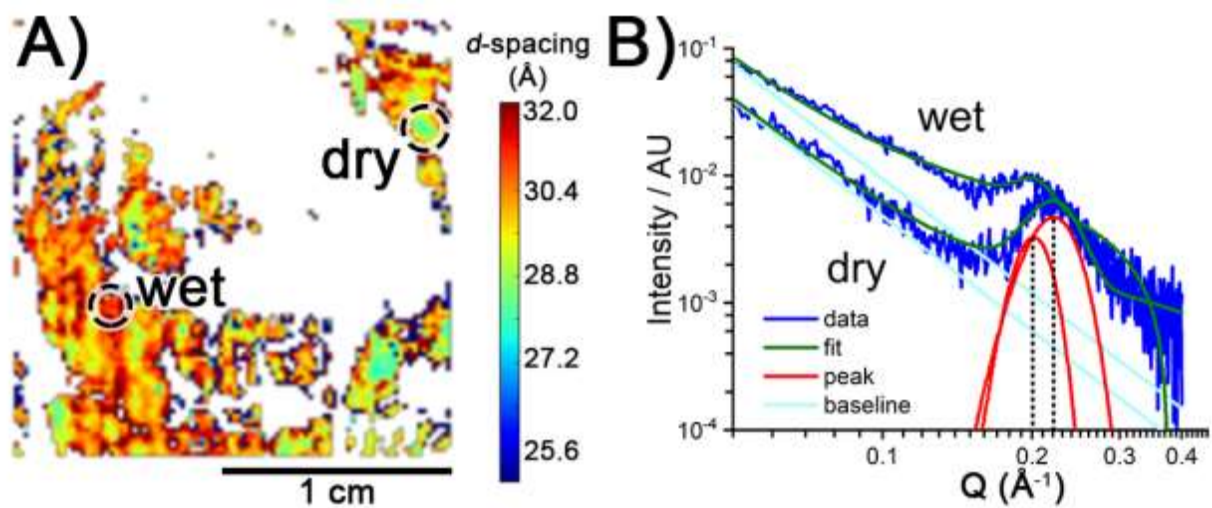

**Fig. 2.** SAXS-CT slice in the plane of the ionomer membrane measured inside the PEMFC operating at 0.8 A/cm², 75°C, 80% relative humidity. A) *d*-spacing of the SAXS ionomer peak, reflecting distance between hydrophilic domains. White regions correspond to locations where the ionomer was not detected in this image slice. B) Reconstructed SAXS patterns extracted from regions marked dry and wet in A. Peaks fitted to the hydrated nanodomain spacing show a shift in the *d*-spacing, reflecting changes in the local polymer hydration.

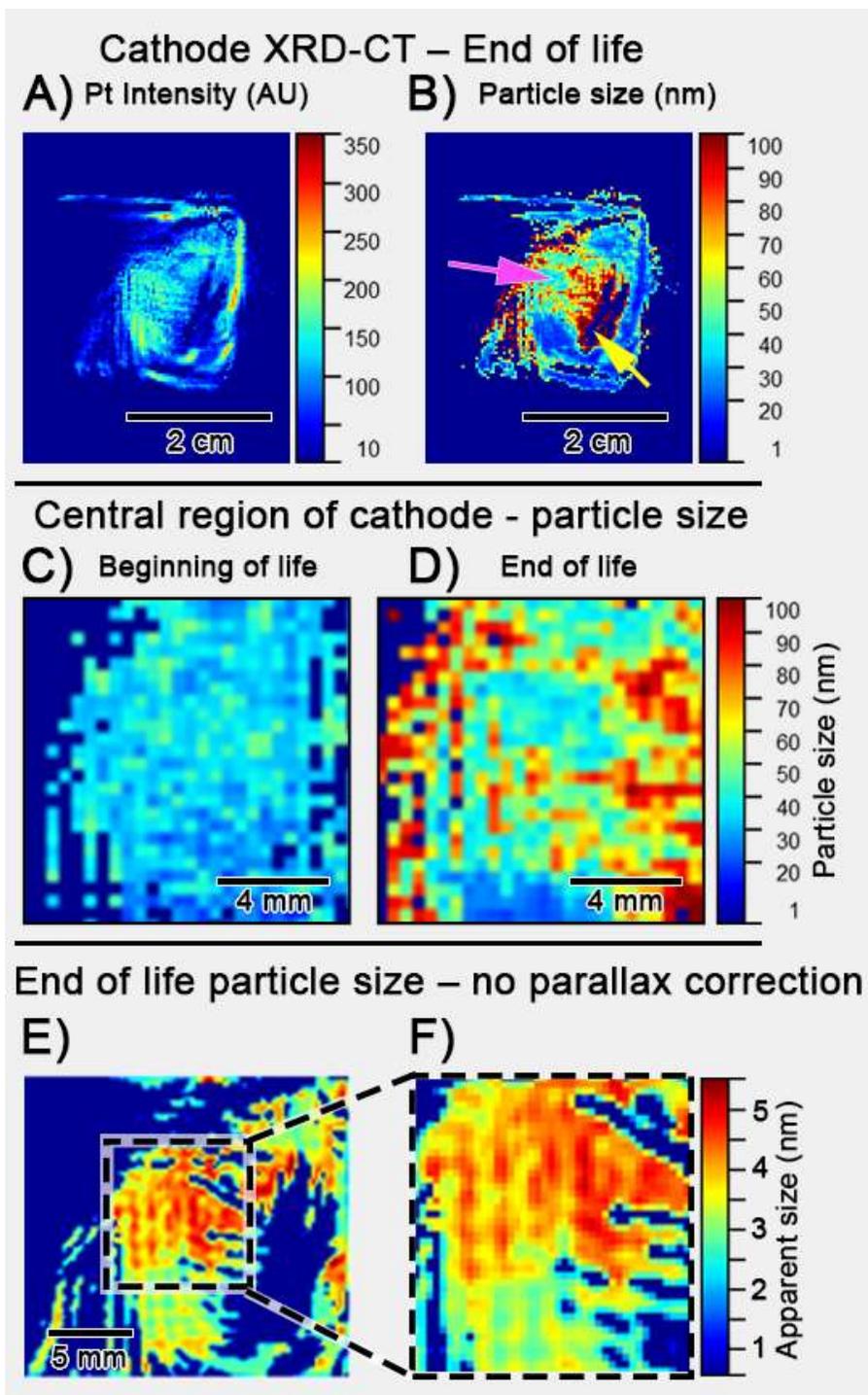

**Fig. 3.** XRD-CT of Pt cathode catalyst. A) Pt intensity distribution over the single CT slice. B) Pt particle size map, parallax corrected.  C, D) Particle size maps for identical regions in the central region of the electrode (magenta arrow in B) at the beginning and end of testing, respectively. E) High resolution crystallite size map for central region of the electrode at end of testing, without parallax correction. F) Magnification of E) showing detailed particle size map.

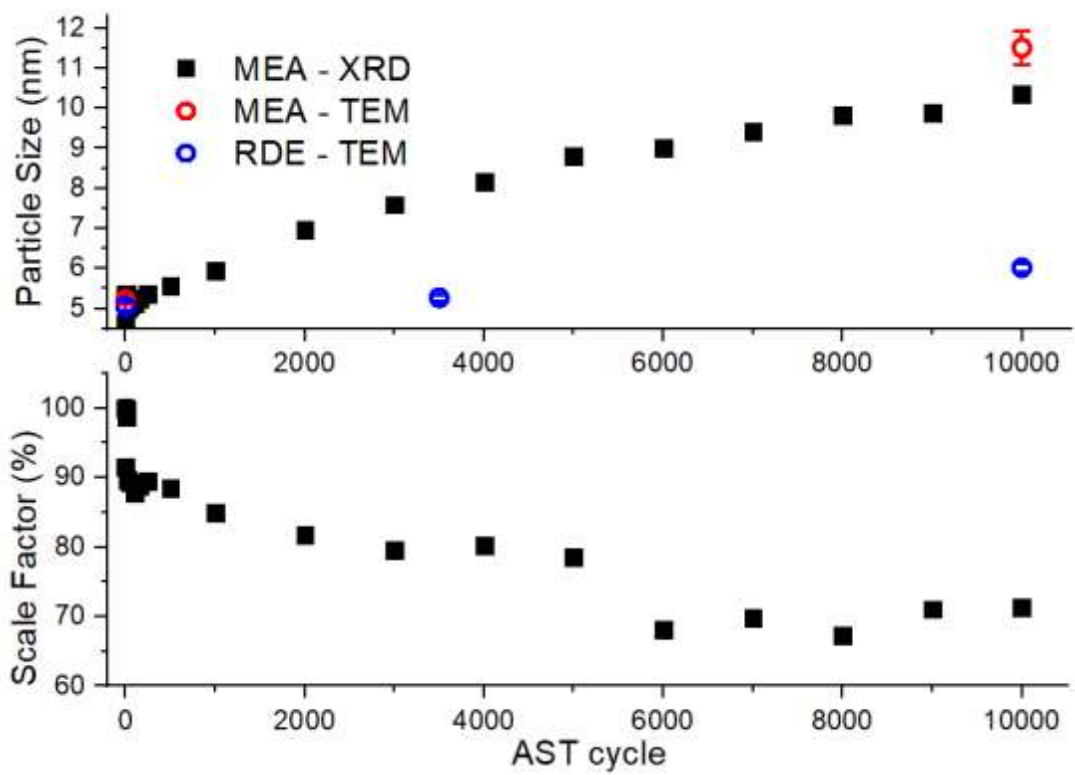

**Fig. 4.** Particle size (top) and percentage of Pt remaining in electrode (bottom) over $10^4$ cycles of accelerated stress testing (0.6-1.0 V, 50 mV/s) determined from Rietveld analysis. Mean particle sizes for the full MEA cathode and rotating disc electrode (RDE) determined by electron microscopy are overlaid (red and blue circles, respectively). The temperature of the PEM and RDE cells during ageing was 80°C and 20°C, respectively. The error estimated from the Rietveld covariance matrix is smaller than the plotted XRD markers. The TEM data error bars correspond to the standard error of the log-normal fit. High resolution crystallite size map for central region of the electrode at end of testing, without parallax correction. F) Magnification of E) showing detailed particle size map.

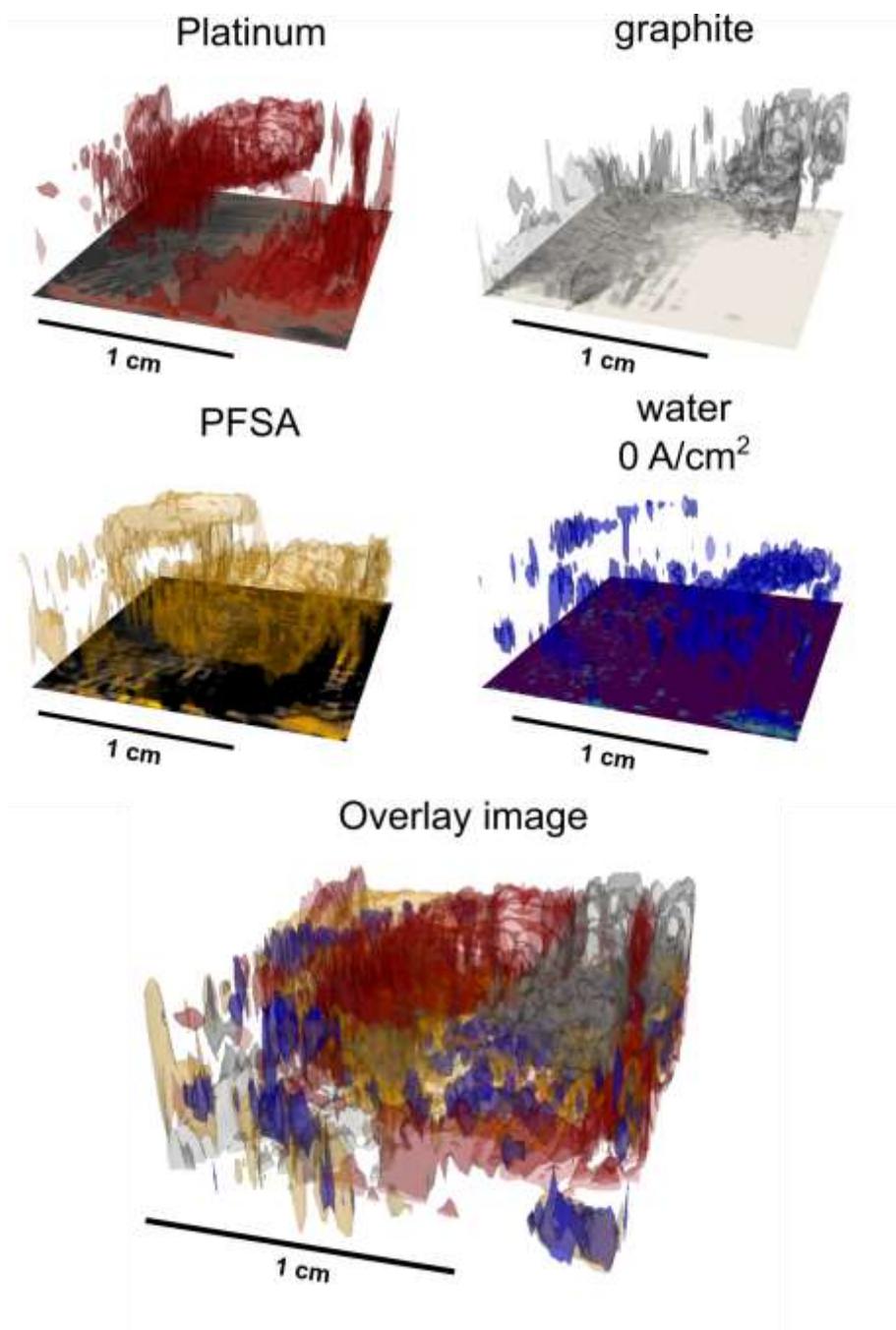

**Fig. S1.** 3D chemical map constructed from the individual components in Fig. 1 in the main article text. An overlay of the Pt, PFSA polymer electrolyte, graphite, and water can be assembled (top), in addition to 3D images of the individual components (bottom).

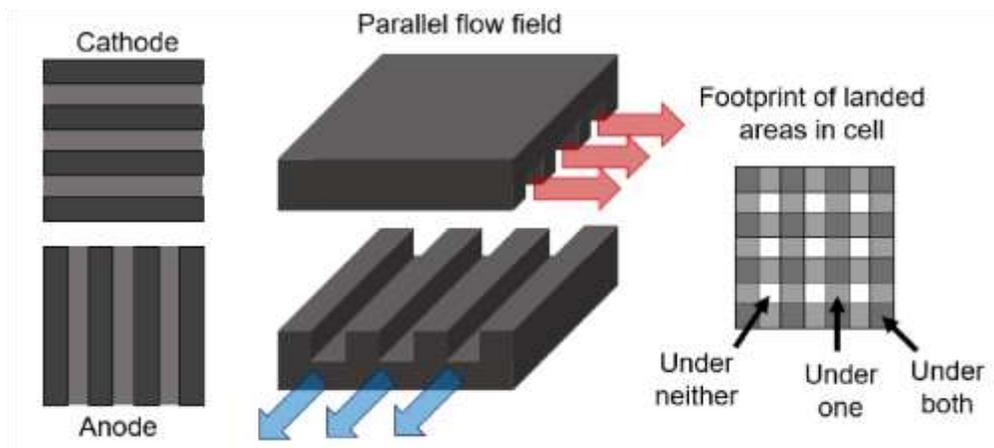

**Fig. S2.** Cartoon of the graphite cathode and anode flow field plates used in the X-ray transparent cell. The ribs of the cathode flow field were oriented perpendicular to the anode flow field for the XRD-CT experiment. The schematic on the right indicates how this pattern creates regions of the MEA which are located under one flow field, both flow fields, or neither. For the SAXS-CT experiment, the ribs of the flow field for anode and cathode were aligned in parallel, overlapping geometry.

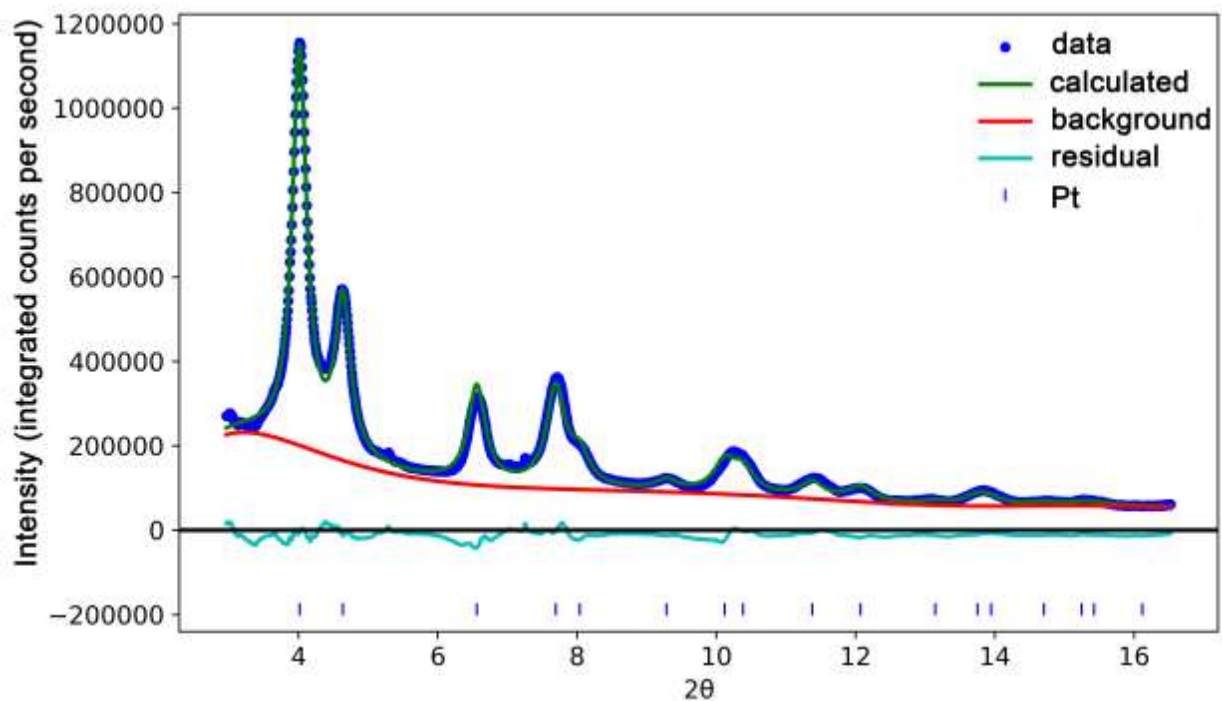

**Fig. S3.** Rietveld refinement of in situ XRD diffractogram from Pt in the cathode catalyst layer, before beginning the accelerated stress test. Blue points represent the observed scattering data, the green line shows the refined model, the red line shows the background polynomial, and the teal line shows the residual. Blue ticks below the curves indicate the calculated positions of the Pt lattice reflections.

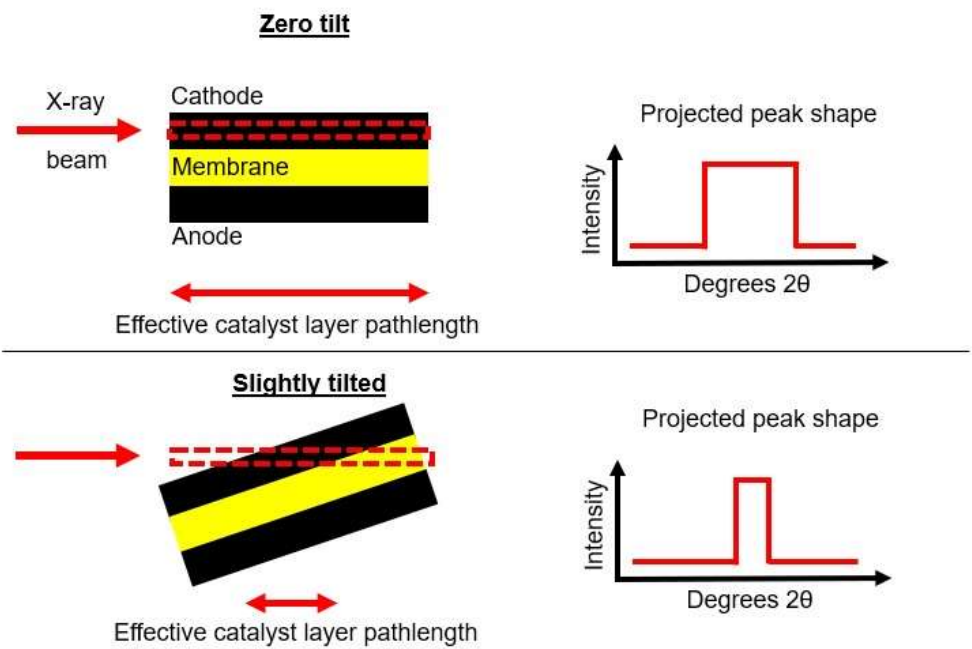

**Fig. S4.** Schematic of the PEM cell during accelerated stress tests, showing the geometry used for collecting Rietveld refinement quality data. Slightly tilting the cell reduces the effective pathlength of the catalyst layer and reduces the parallax effect. The cell was tilted only 0.3° and has been exaggerated here for clarity.

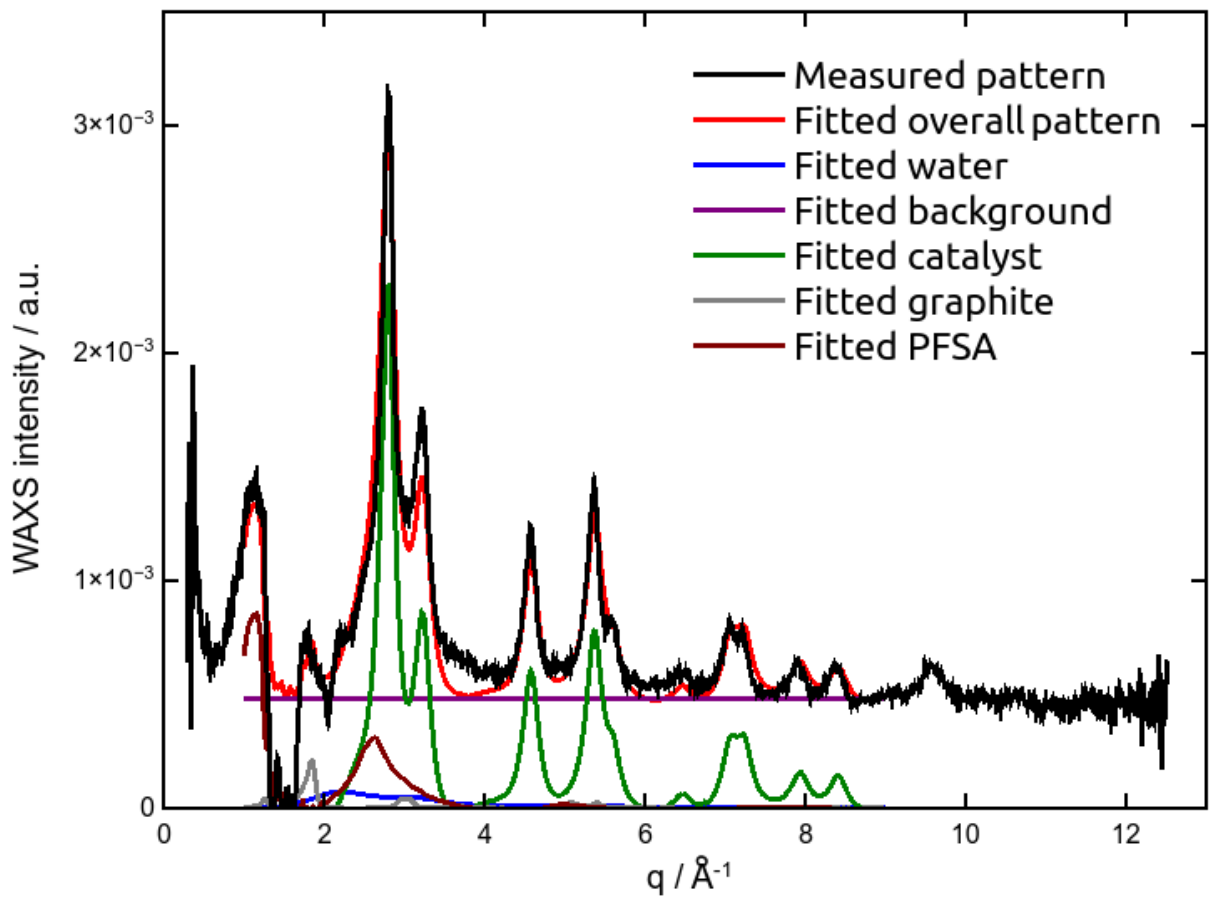

**Fig. S5.** Deconvolution of each phase from a single reconstructed XRD-CT voxel.

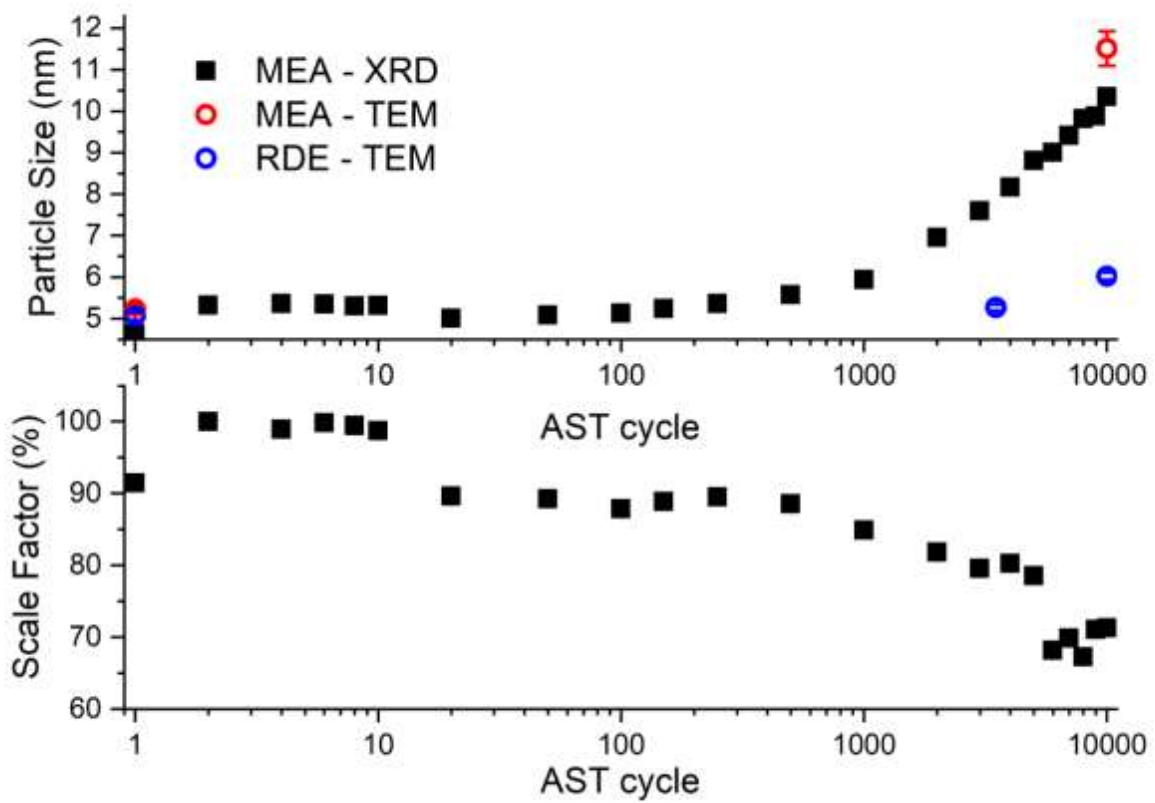

**Fig. S6.** Accelerated stress testing plot identical to Fig. 4 in the main article text, but with logarithmic scale for number of cycles.

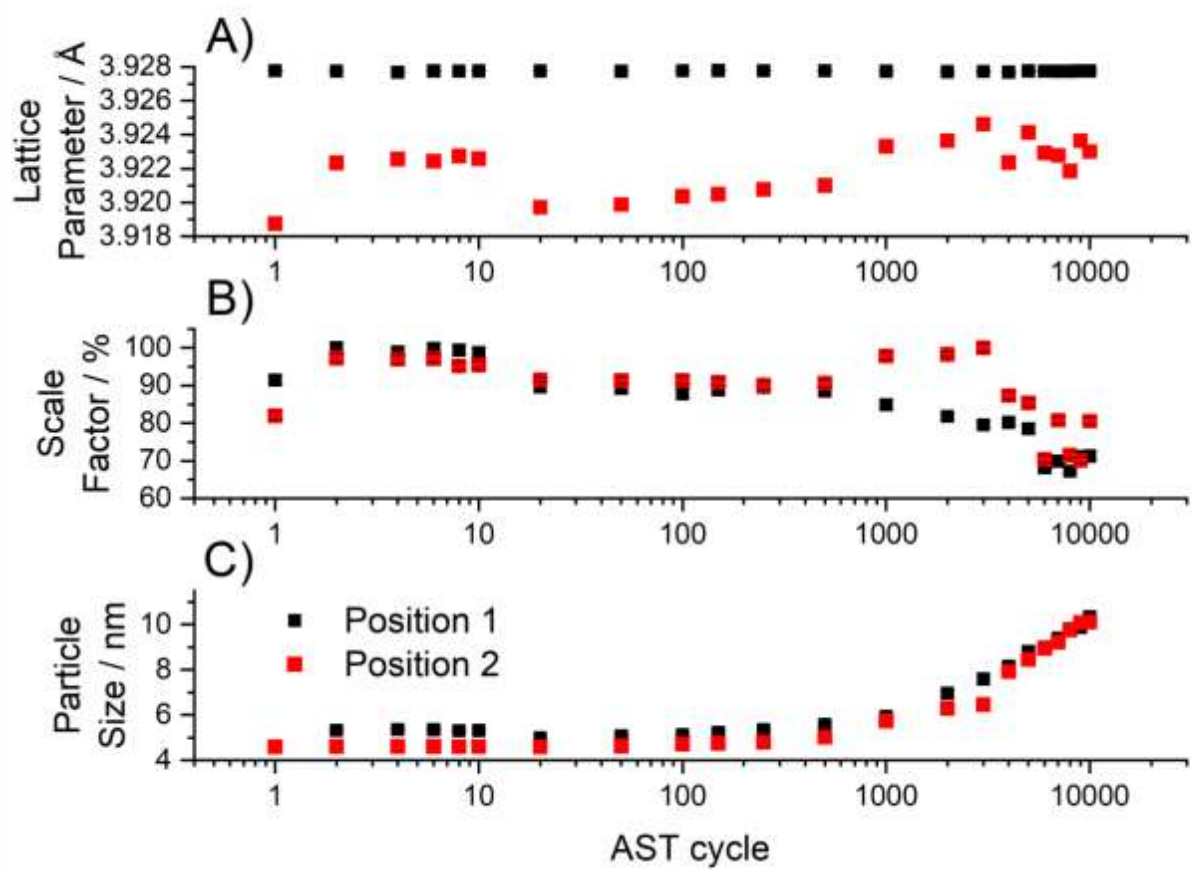

**Fig. S7.** Comparison of landed and channel areas with *in situ* diffraction during accelerated stress testing under inert atmosphere. Position 1 was measured down the length of the flow field's landed area (black squares). Position 2 was measured in the adjacent channel of the flow field (red squares). Rietveld analysis of the Pt phase from the two locations yielded the lattice parameter, scale factor, and particle size (A, B, and C respectively).

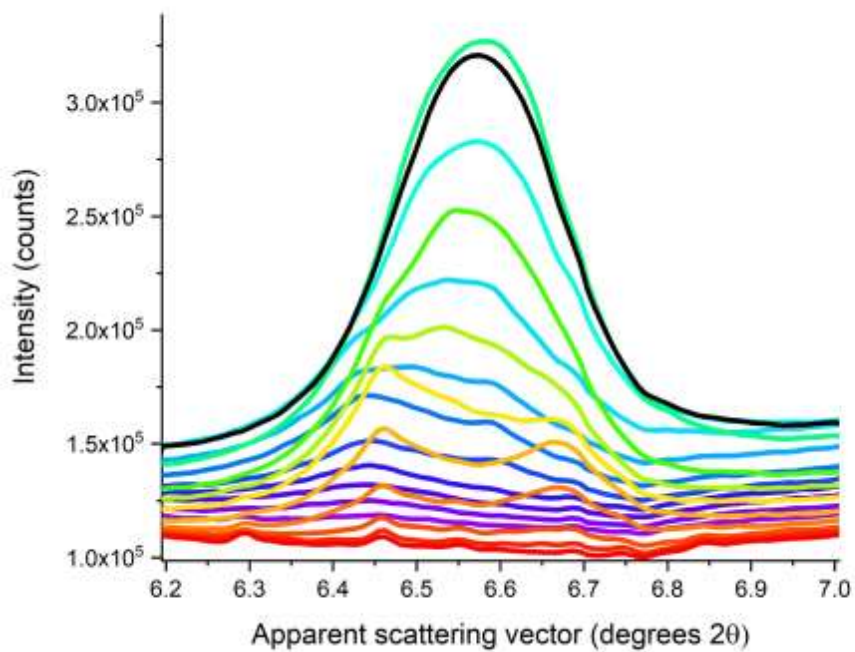

**Fig. S8.** Pt 220 reflection while scanning the beam vertically through the cross-section of the catalyst layer. Diffractograms were collected in 10 μm steps. The position aligned with the catalyst layer (black trace) has a smooth, symmetric Lorentzian shape and good intensity. Beam positions too high or too low suffer from progressively worse peak broadening/splitting.

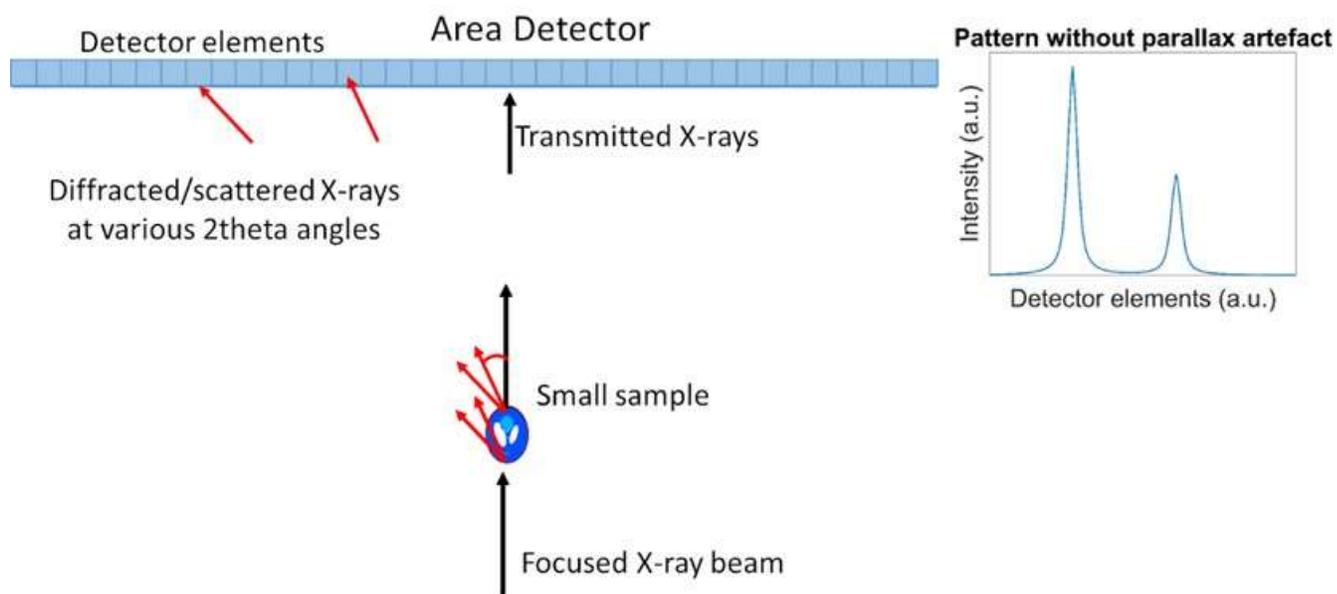

**Fig. S9.** Illustration of a typical diffraction measurement in transmission geometry with a focused X-ray beam and an area detector. The X-rays scattered/diffracted along the sample at a certain 2θ angle arrive at the same detector element.

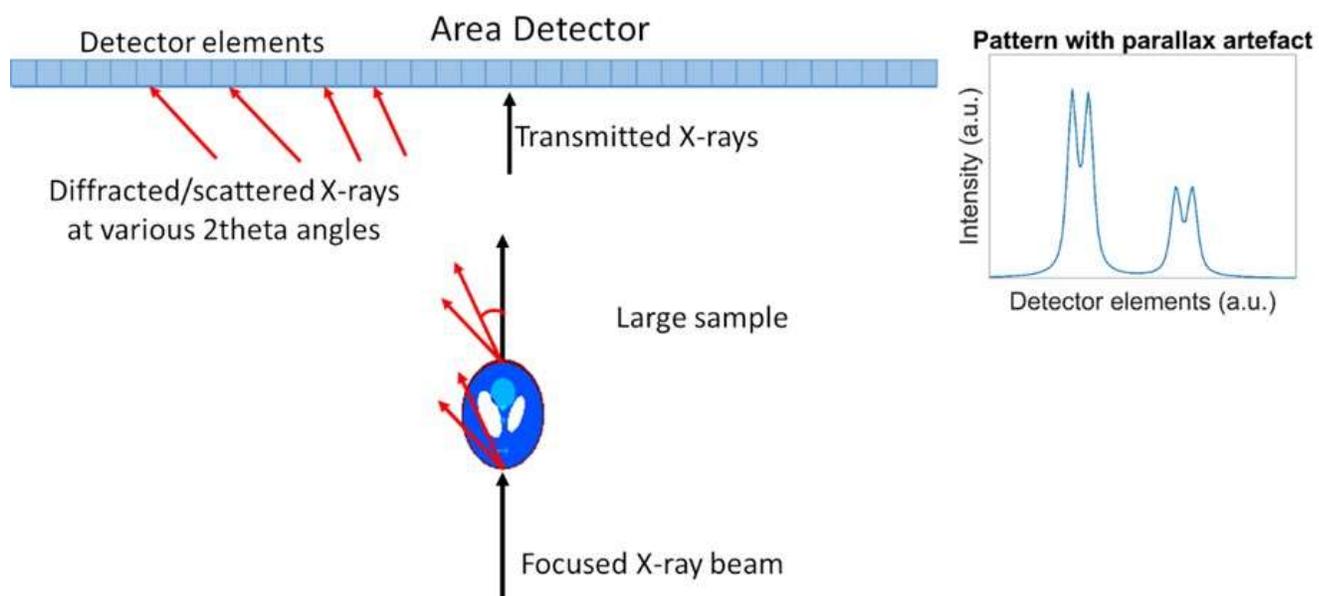

**Fig. S10.** Illustration of a diffraction measurement in transmission geometry with parallax artefact. The X-rays scattered/diffracted along the sample at a certain 2θ angle arrive at different detector elements leading to peak broadening, and even peak splitting.

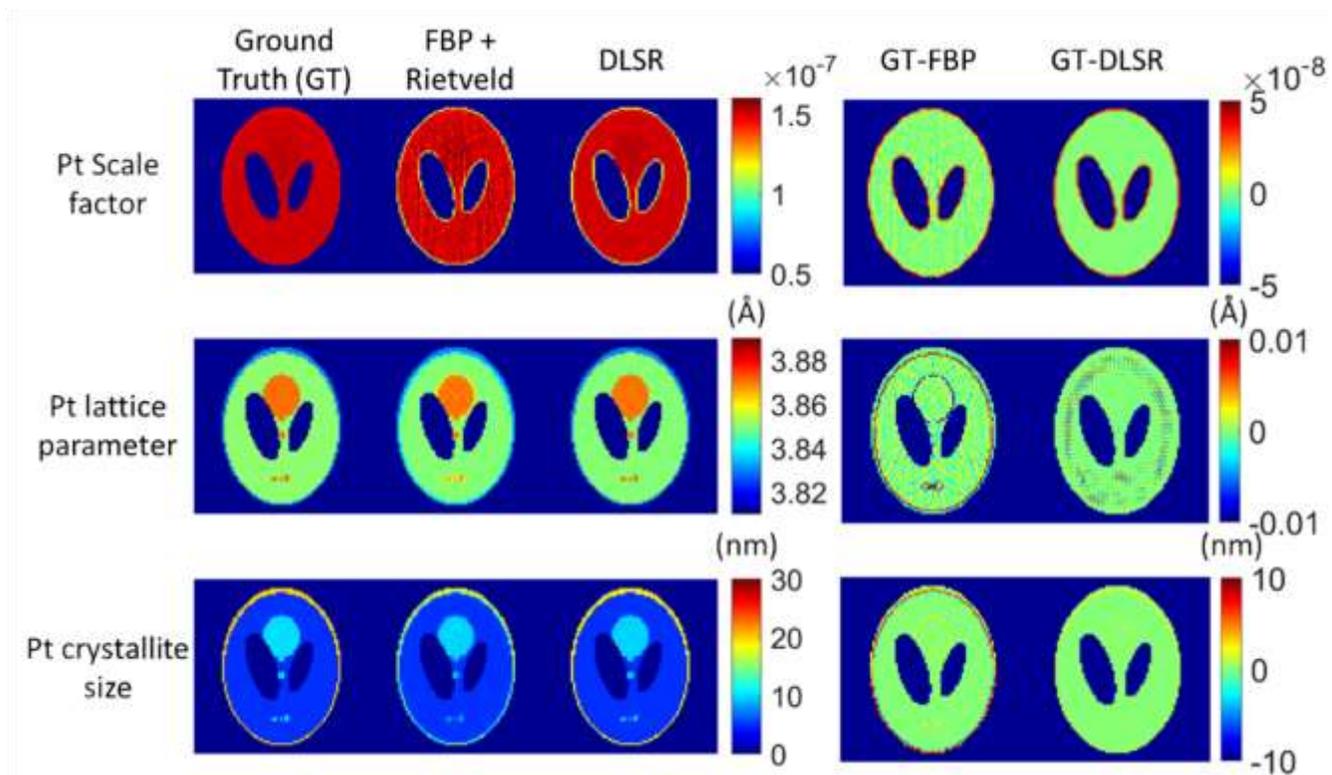

**Fig. S11.** Left: Simulated parallax-free Pt XRD-CT dataset (Ground Truth) and results obtained using the reverse analysis FBP method and the DLSR algorithm. Right: Difference maps between the Ground Truth images and the ones obtained from the two analysis methods.

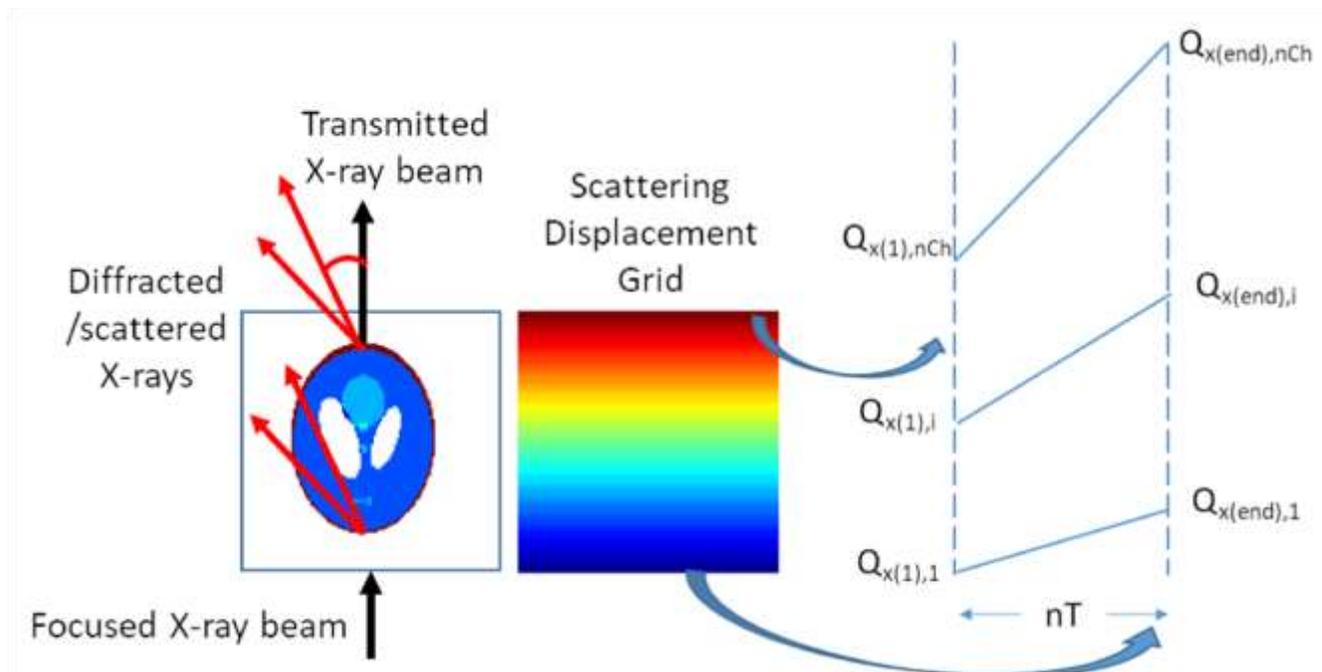

**Fig. S12.** The Scattering Displacement Grid corresponds to a 3D matrix that has (nT, nT, nCh) dimensions; each voxel corresponds to a unique scattering axis (2θ/d/Q).

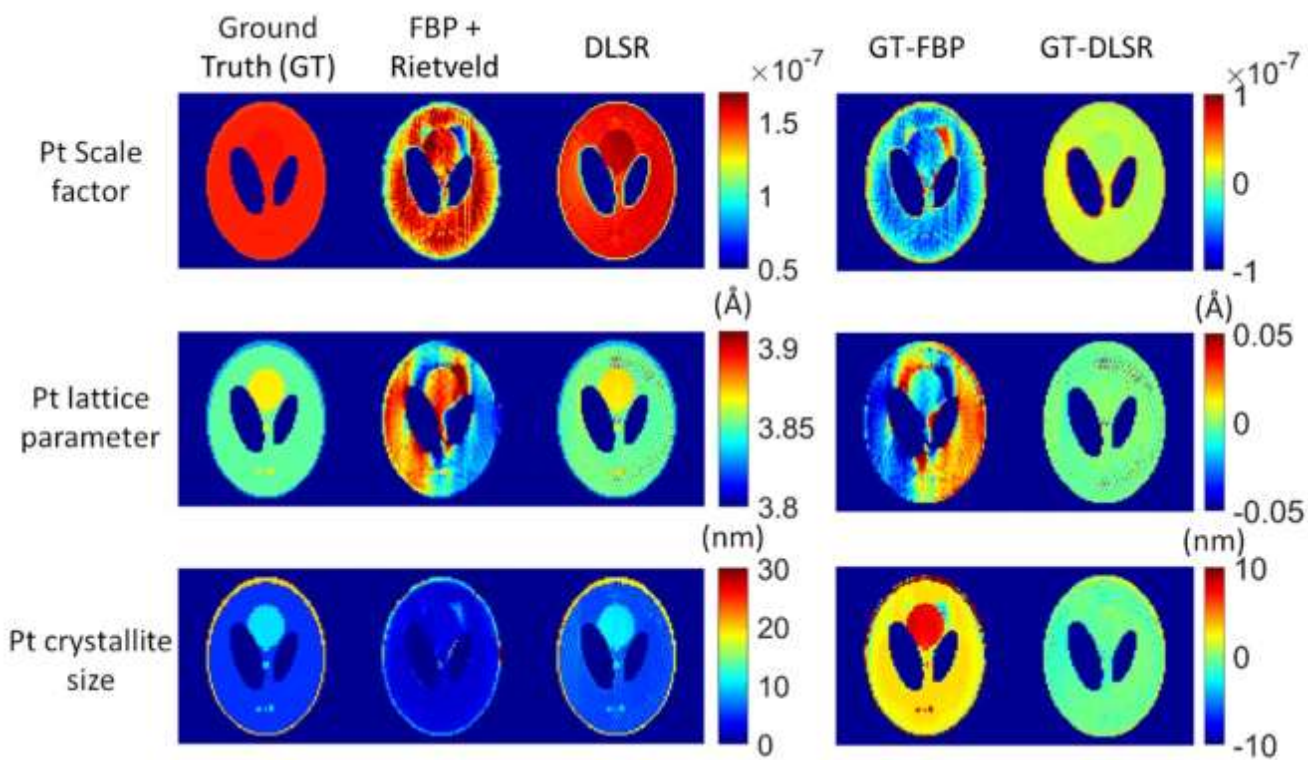

**Fig. S13.** Left: Simulated parallax Pt XRD-CT dataset (Ground Truth) and results obtained using the reverse analysis method and the DLSR algorithm. Right: Difference maps between the Ground Truth images and the ones obtained from the two analysis methods.

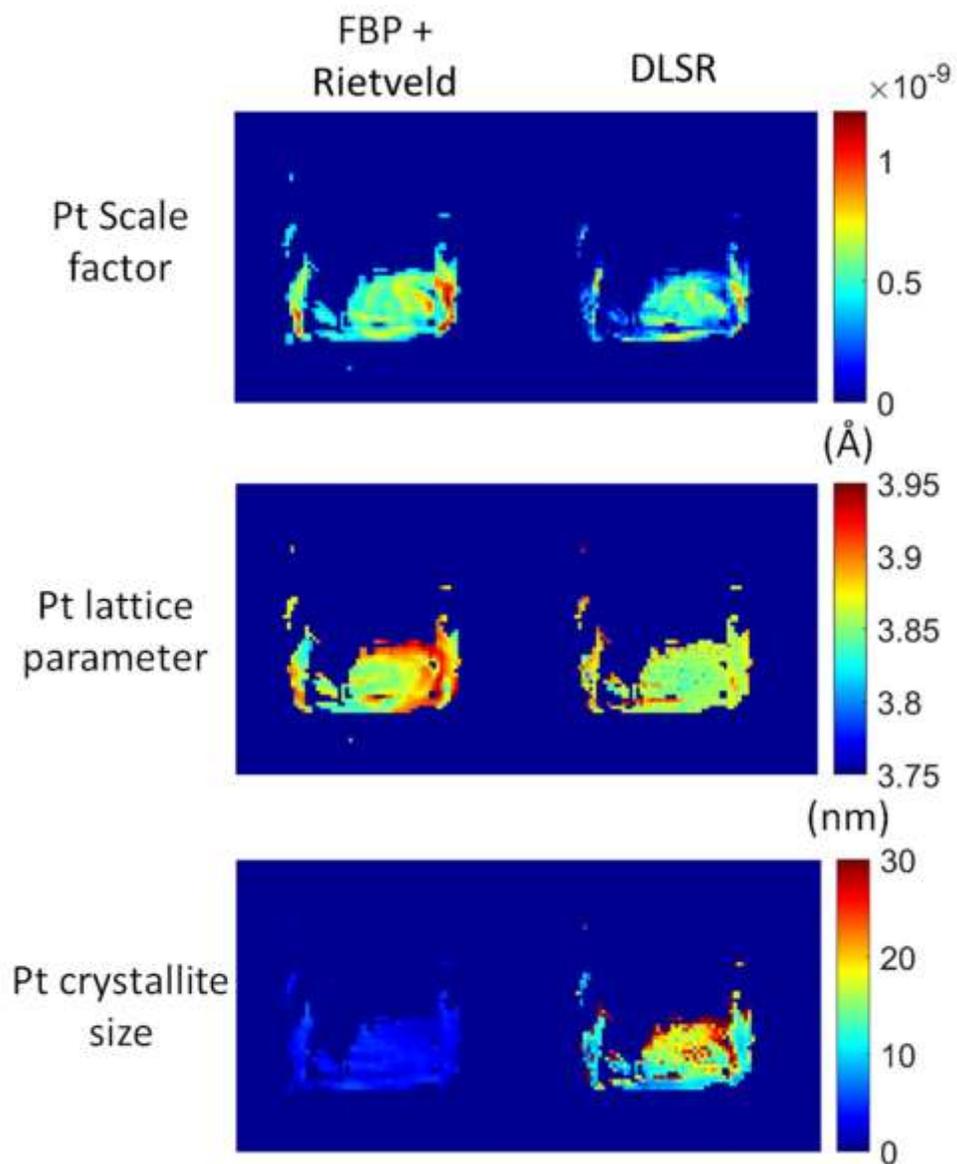

**Fig. S14.** Reconstructions obtained from a single slice of the PEMFC XRD-CT dataset containing parallax artefact using the reverse analysis method and the DLSR algorithm.

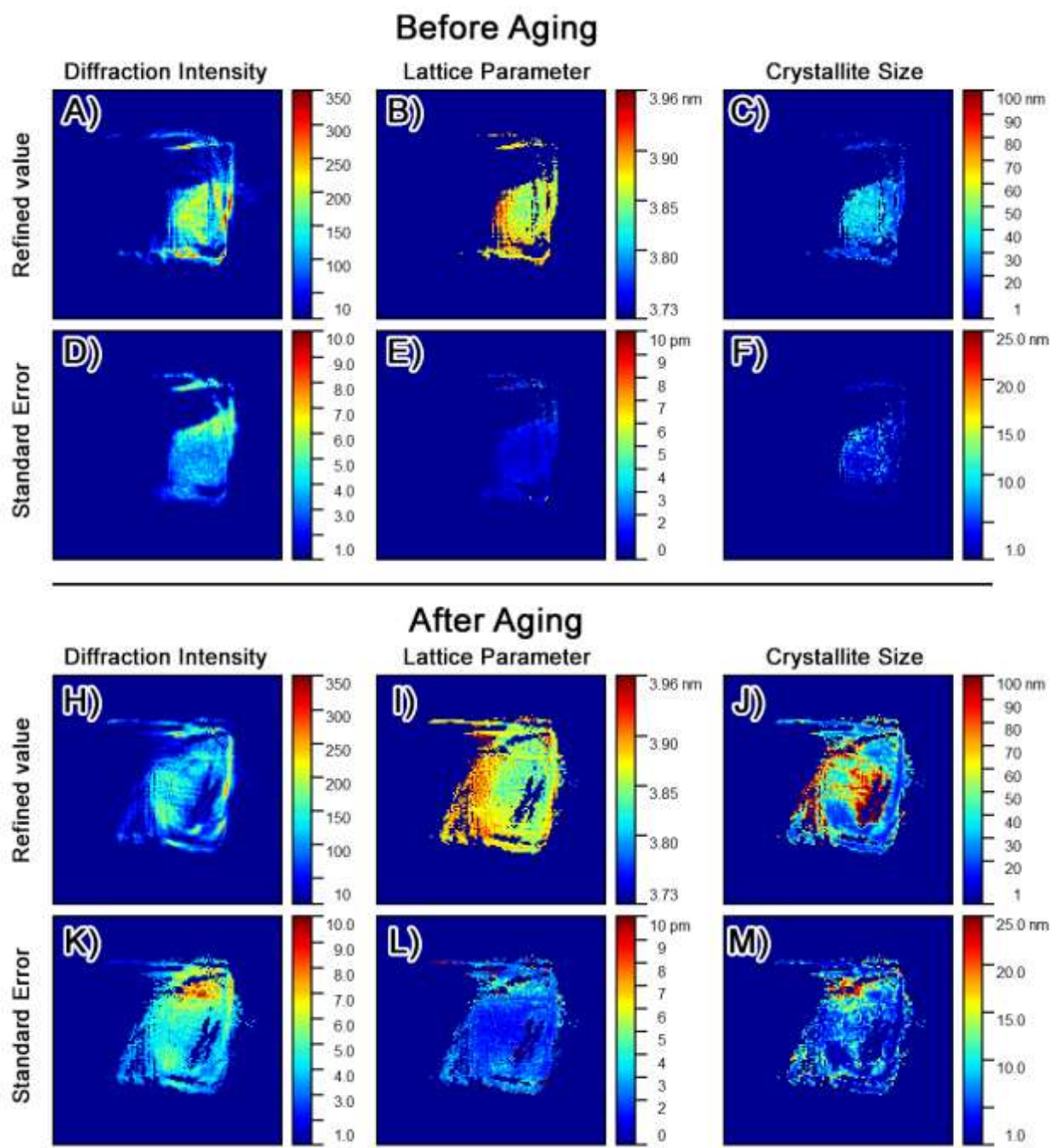

**Fig. S15.** XRD-CT slices through the cathode at before and after degradation, with parallax correction. The map of refined values, along with the estimated error is presented for the diffraction intensity, lattice parameter, and crystallite size. These images correspond to the maps shown in Figs. 3A-3D.

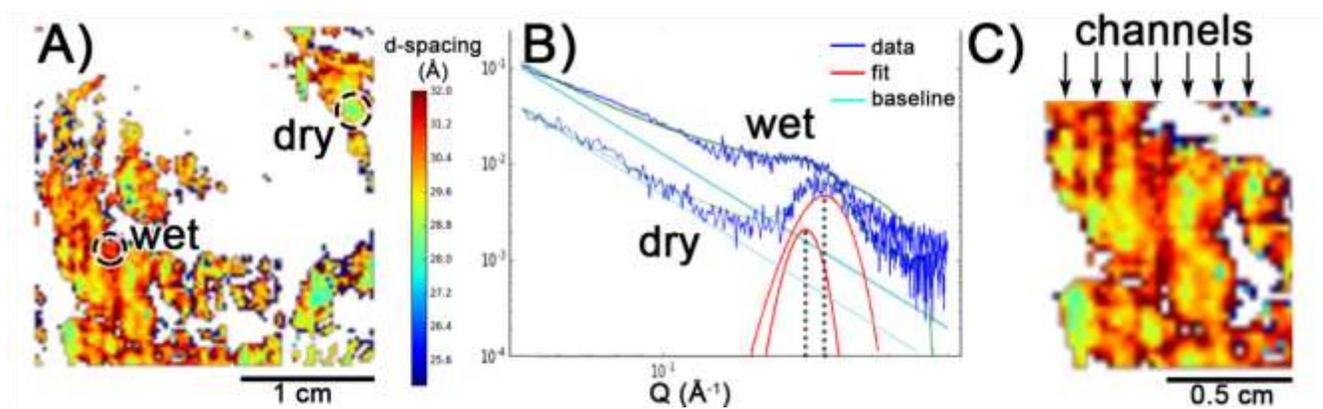

**Fig. S15.** SAXS-CT image identical to Fig. 2, but with detailed panel C) highlighting the formation of striped, hydrated regions in the membrane.

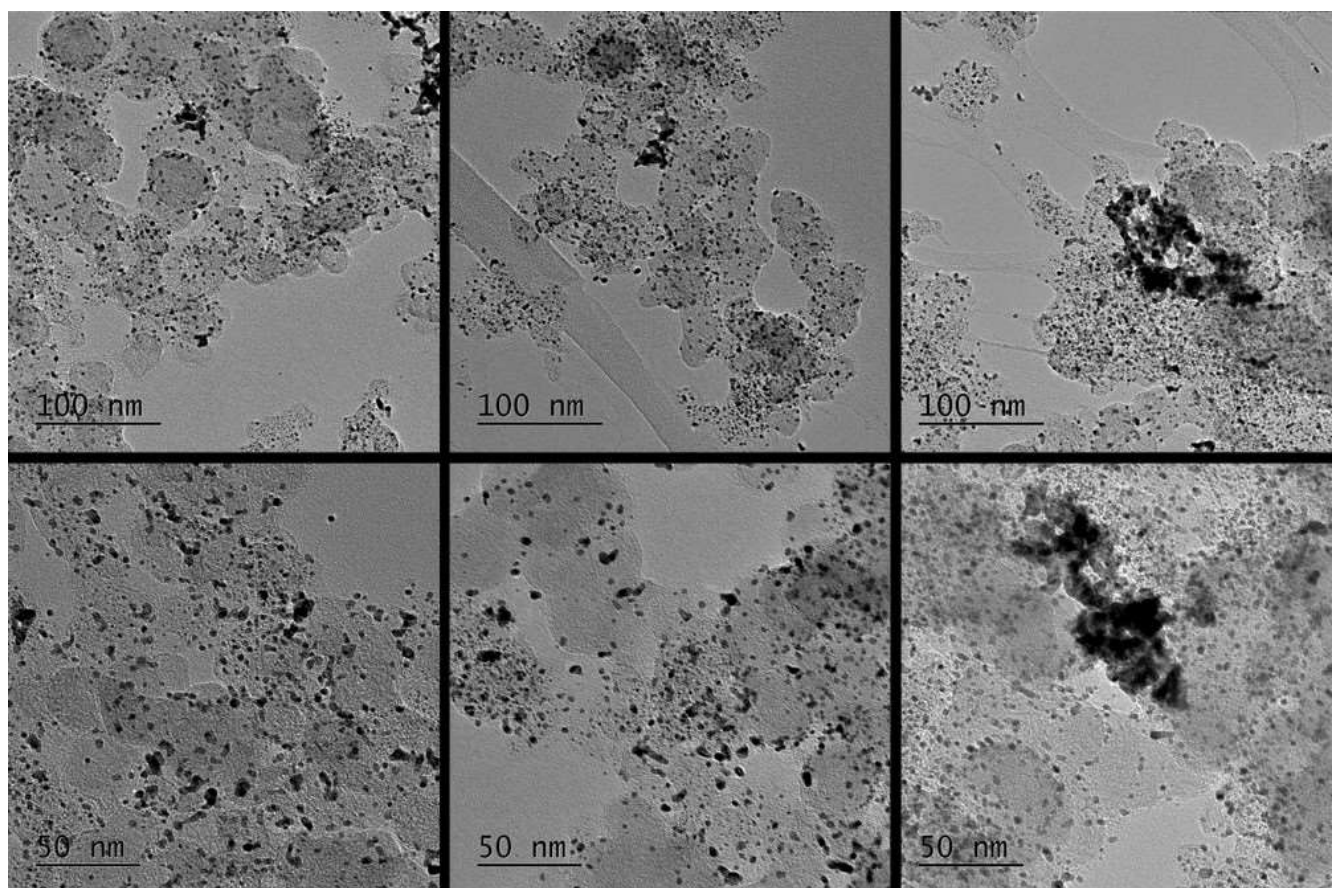

**Fig. S17.** TEM images of the fresh catalyst used in the MEA for accelerated stress testing.

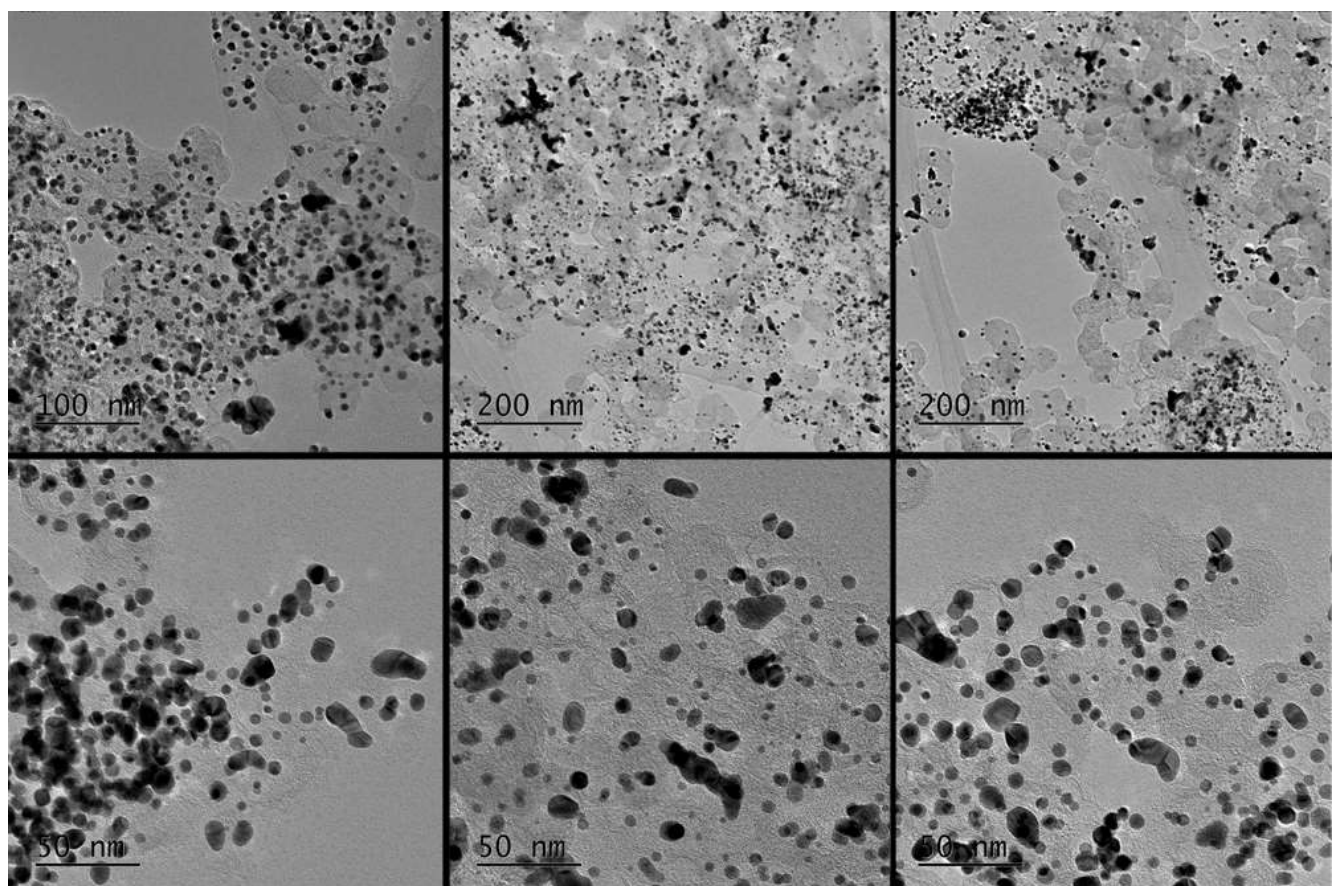

**Fig. S18.** TEM images of the catalyst used in the MEA for accelerated stress testing, after 10000 cycles.

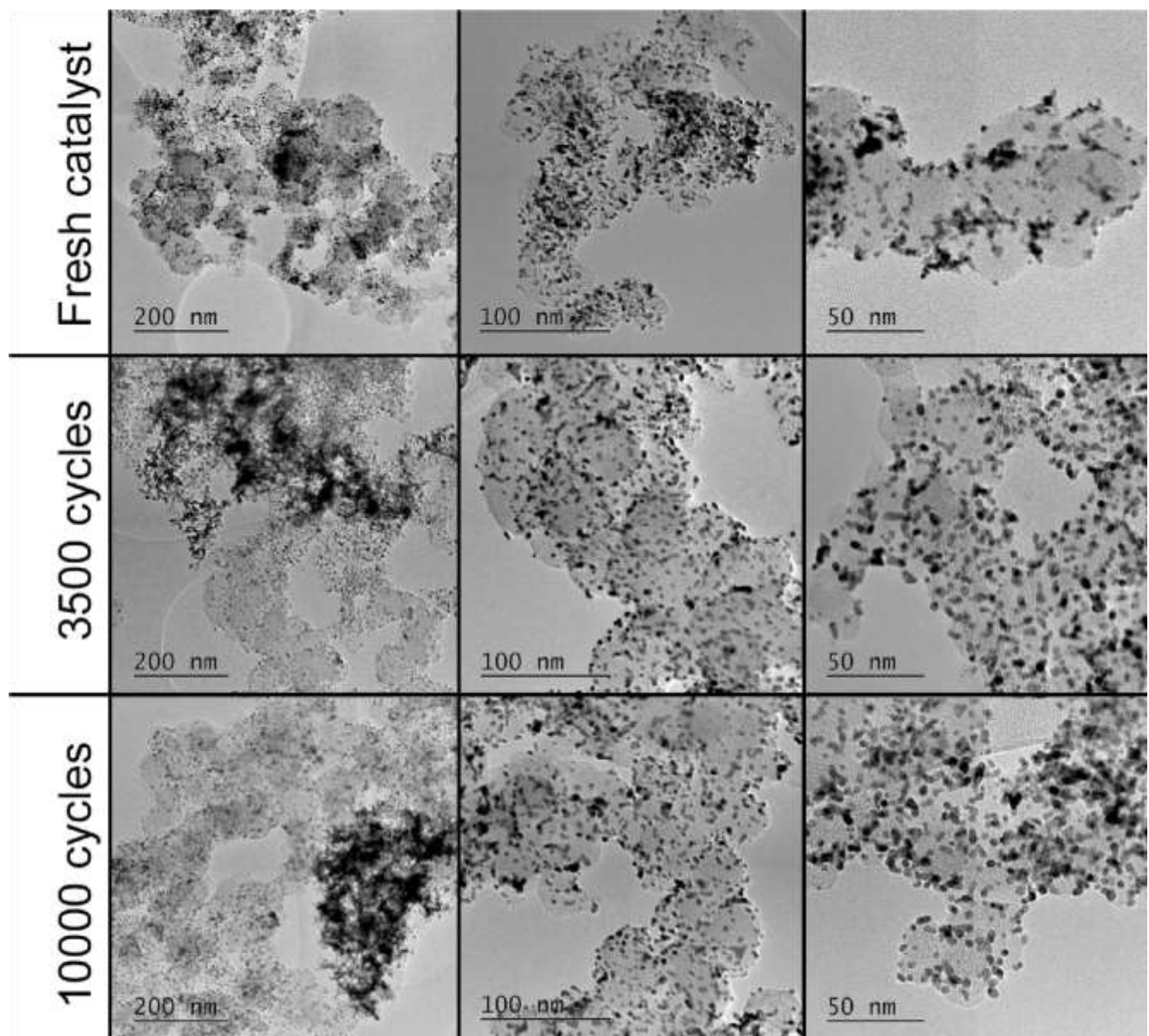

**Fig. S19.** TEM images of the catalyst used in the RDE cell for accelerated stress testing. The top, middle, and bottom rows correspond to the fresh catalyst, and samples collected off the RDE tip after 3500 and 10000 cycles.

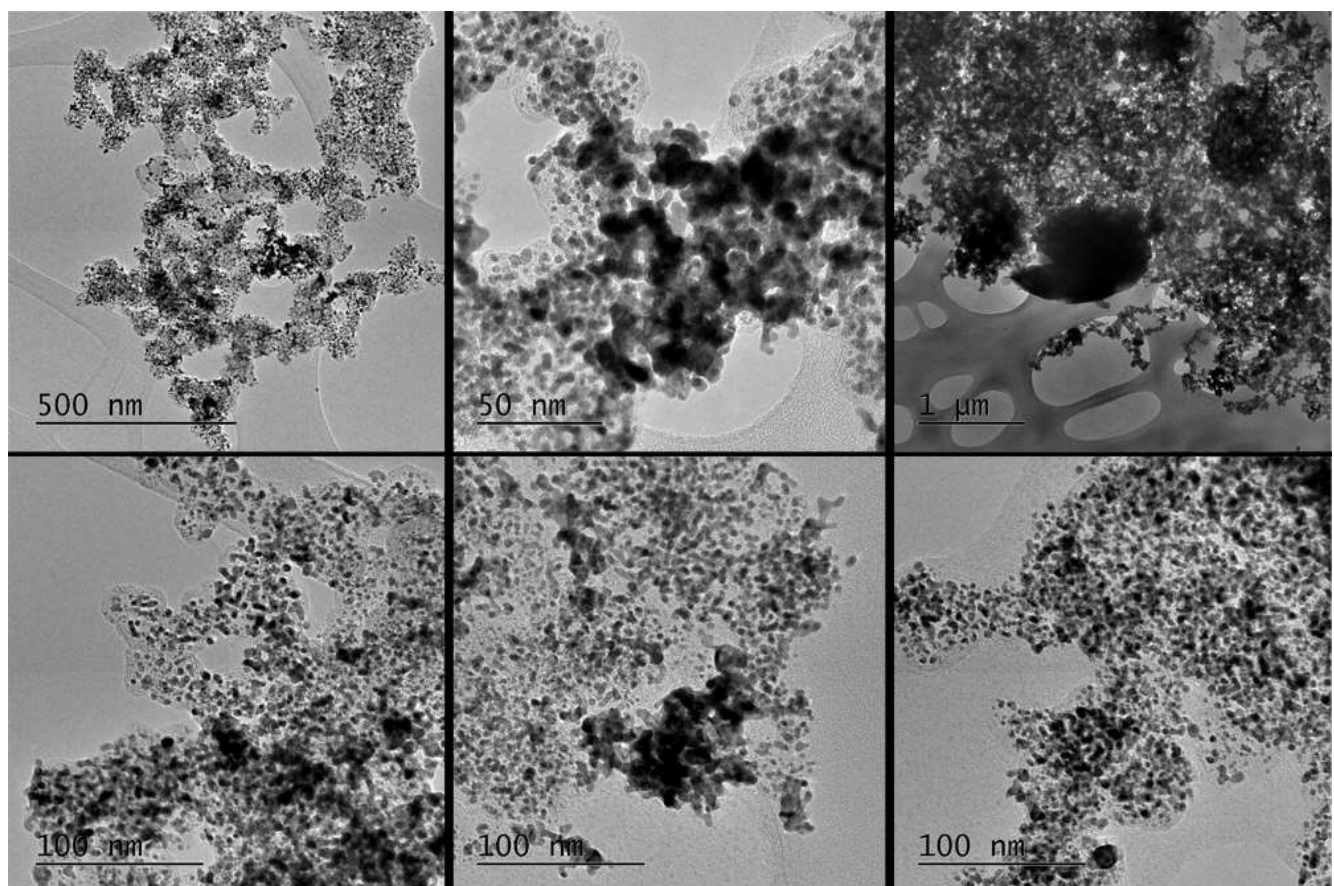

**Fig. S2016.** TEM images of the catalyst used in the MEA for the XRD-CT experiment. Note the presence of large nanoparticles and aggregates not found in the other samples.

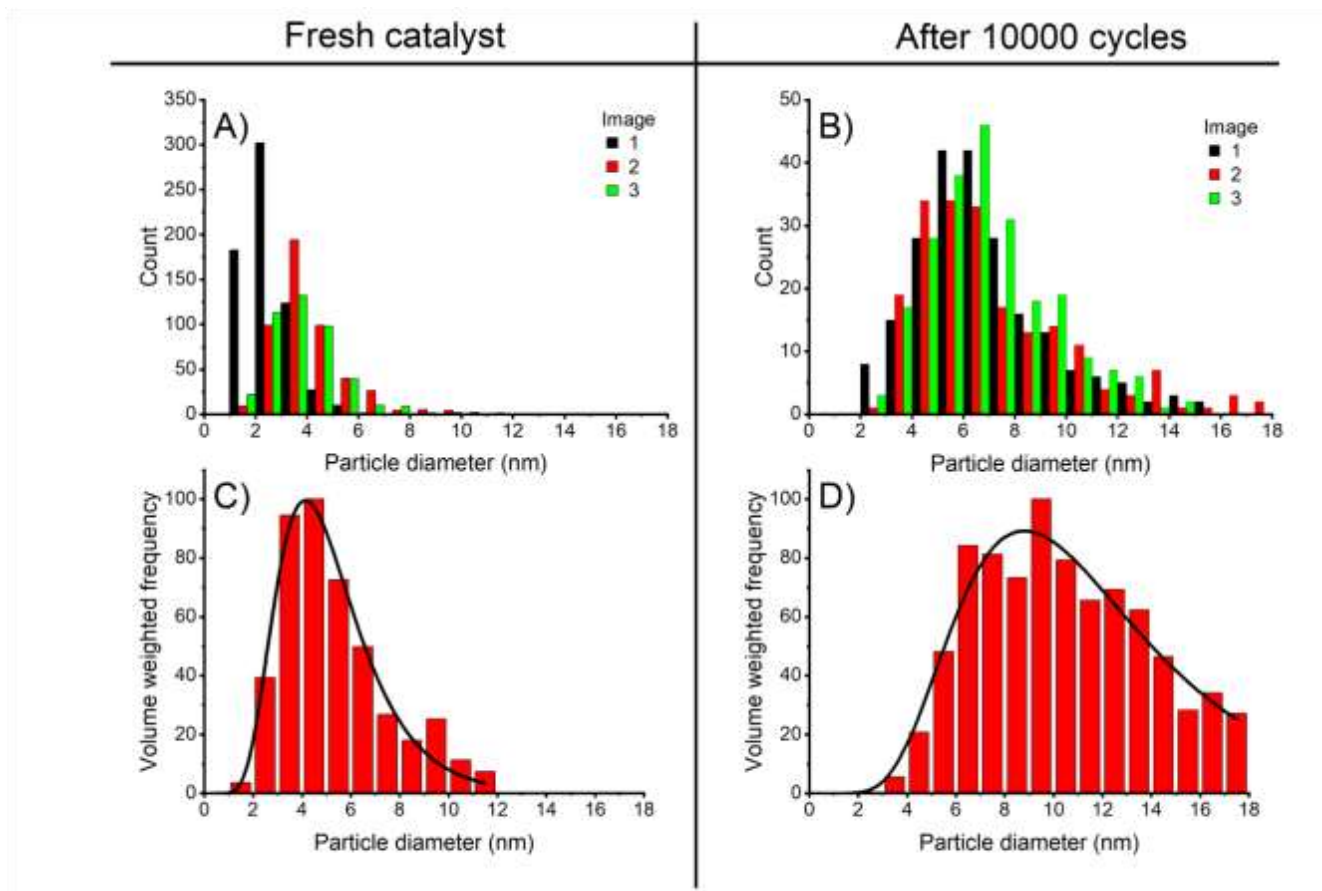

**Fig. S21.** TEM particle size distribution of the catalyst used in the MEA for accelerated stress testing. A & B) Histogram of particle sizes from three TEM images of fresh catalyst and after 10000 cycles, respectively. The black, red, and green bars correspond to the particles imaged in different locations across the TEM grid. C & D) Volume weighted, normalized histograms of the sum of all three images for fresh and aged catalyst, respectively. The black lines represent log-normal distribution fits. 1556 and 639 particles were counted for the fresh and aged catalysts, respectively.

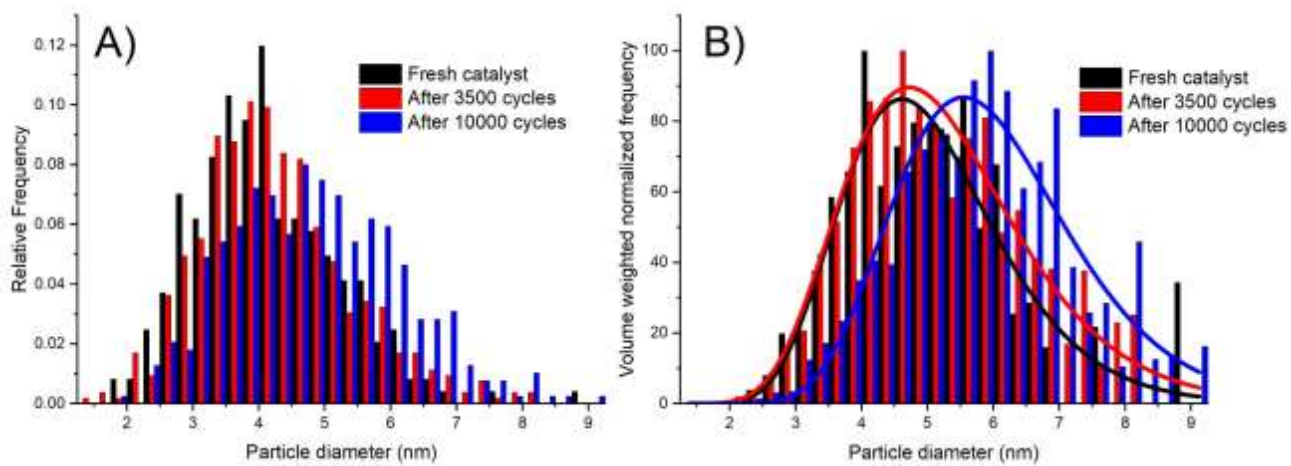

**Fig. S172.** TEM particle size distribution of the catalyst used in the RDE cell for accelerated stress testing. Black, red, and blue bars respectively correspond to the distributions at the beginning of the test, after 3500 cycles, and after 10000 cycles. A) The number-weighted histograms of the three samples. B) The volume weighted, normalized distributions of the same data. Solid lines represent log-normal distribution fits to each sample. 242, 524, and 387 particles were counted for the catalysts at the beginning of the test, after 3500 cycles, and after 10000 cycles, respectively.

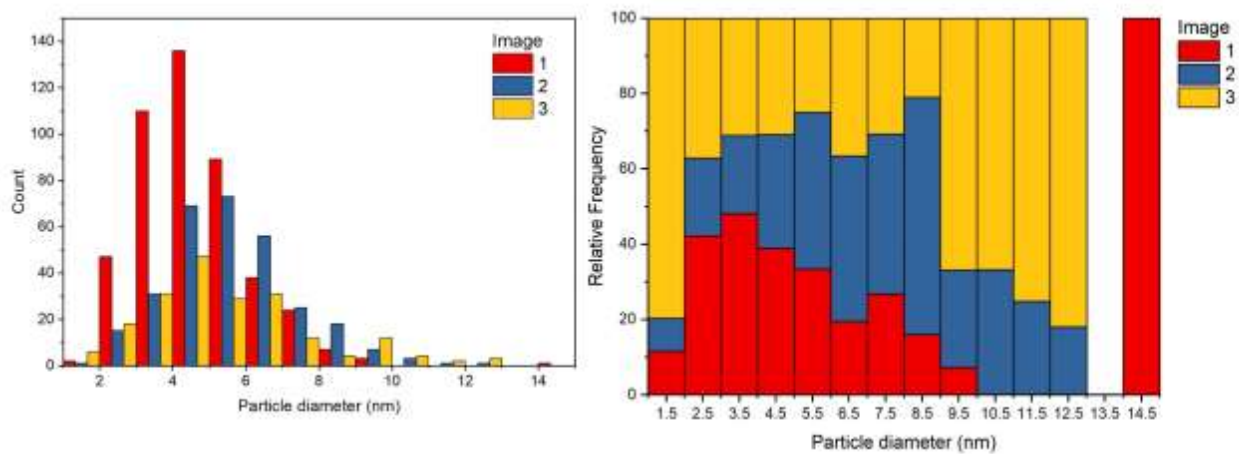

**Fig. S23.** TEM particle size distribution of the fresh catalyst used in the MEA for XRD-CT. Number weighted histogram of particles from three TEM images (left). Relative frequency for each size bin between the three images (right) show the heterogeneity in size distribution. 457 particles were counted.

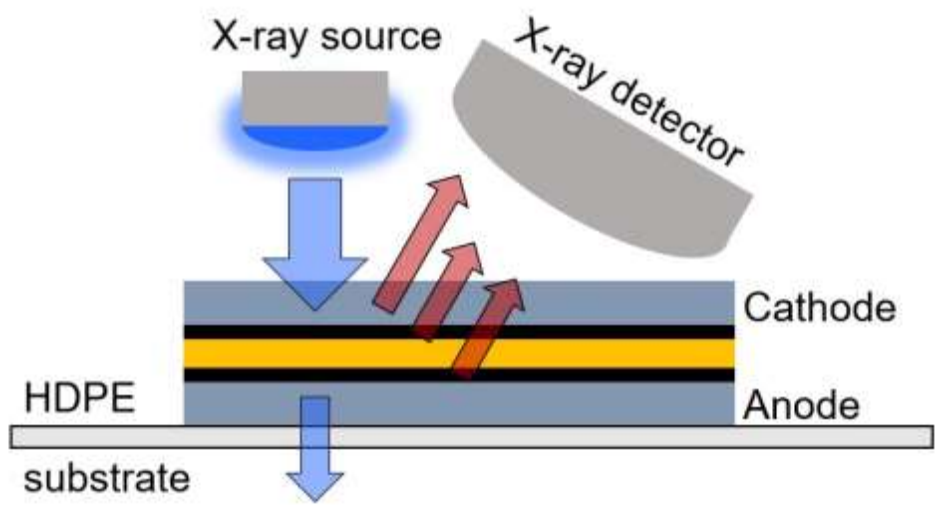

**Fig. S18.** Schematic of the through-plane XRF microscopy experiment.

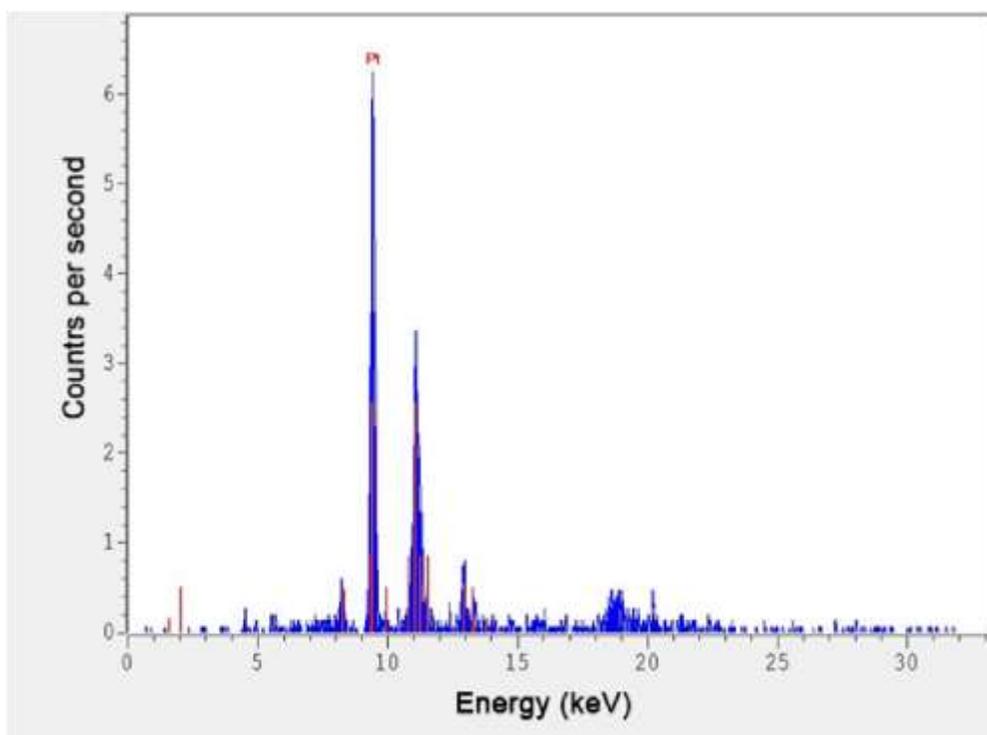

**Fig. S195.** XRF spectra collected from the central region of the MEA. Red lines indicate the positions and calculated intensities for a Pt reference.

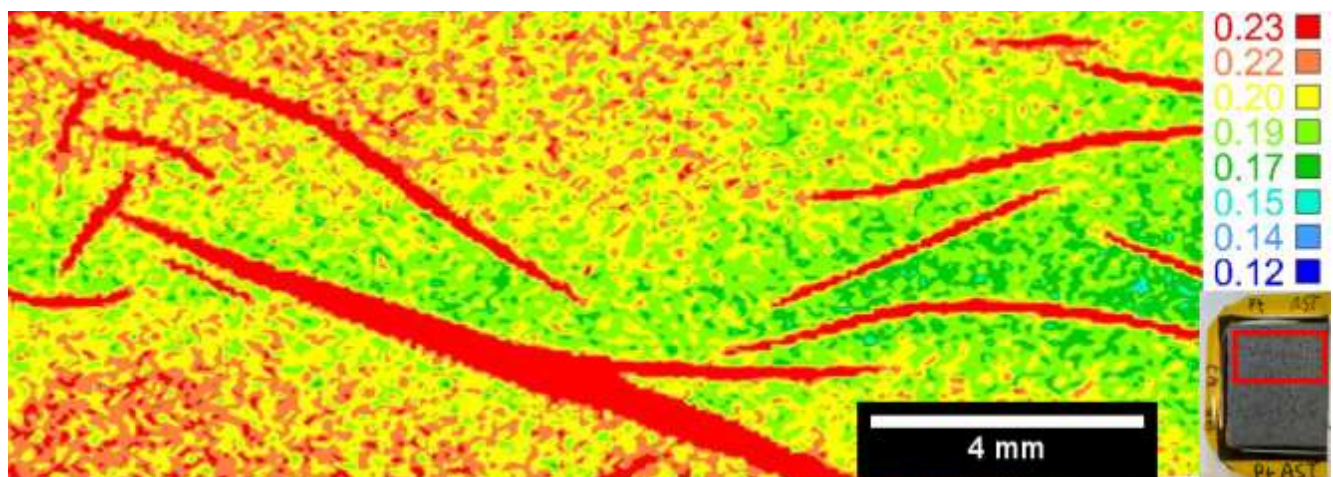

**Fig. S206.** XRF heat map of the MEA after being subjected to accelerated stress testing. Inset: image of the 5 cm$^2$ MEA sample with red rectangle indicating the measured area. Numbers correspond to the intensity of summed Pt band intensities, in arbitrary units.

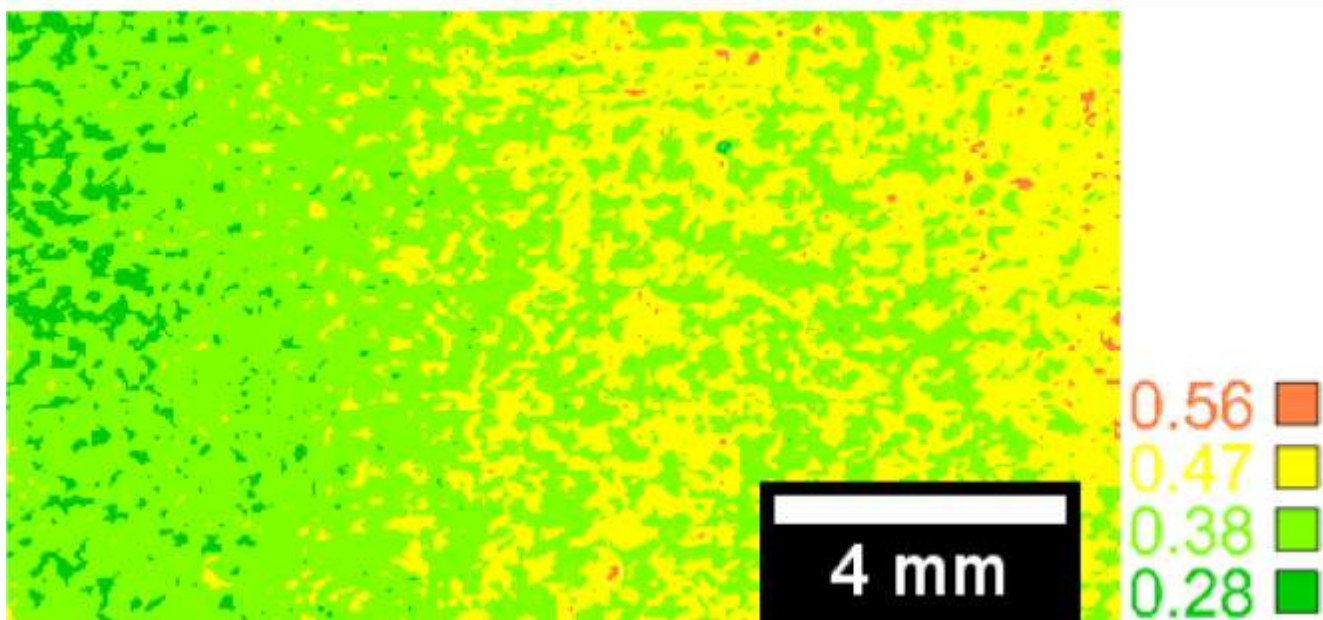

**Fig. 27.** XRF heat map of the MEA used for XRD-CT imaging. Numbers correspond to the intensity of summed Pt band intensities, in arbitrary units.

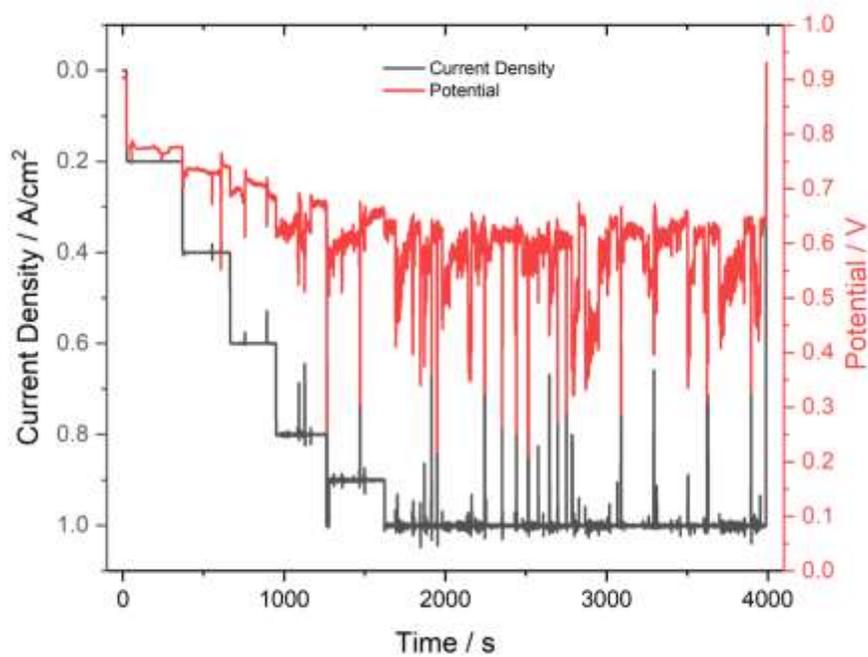

**Fig. S218.** Initial conditioning of the fresh MEA before polarization curves and accelerated stress testing.

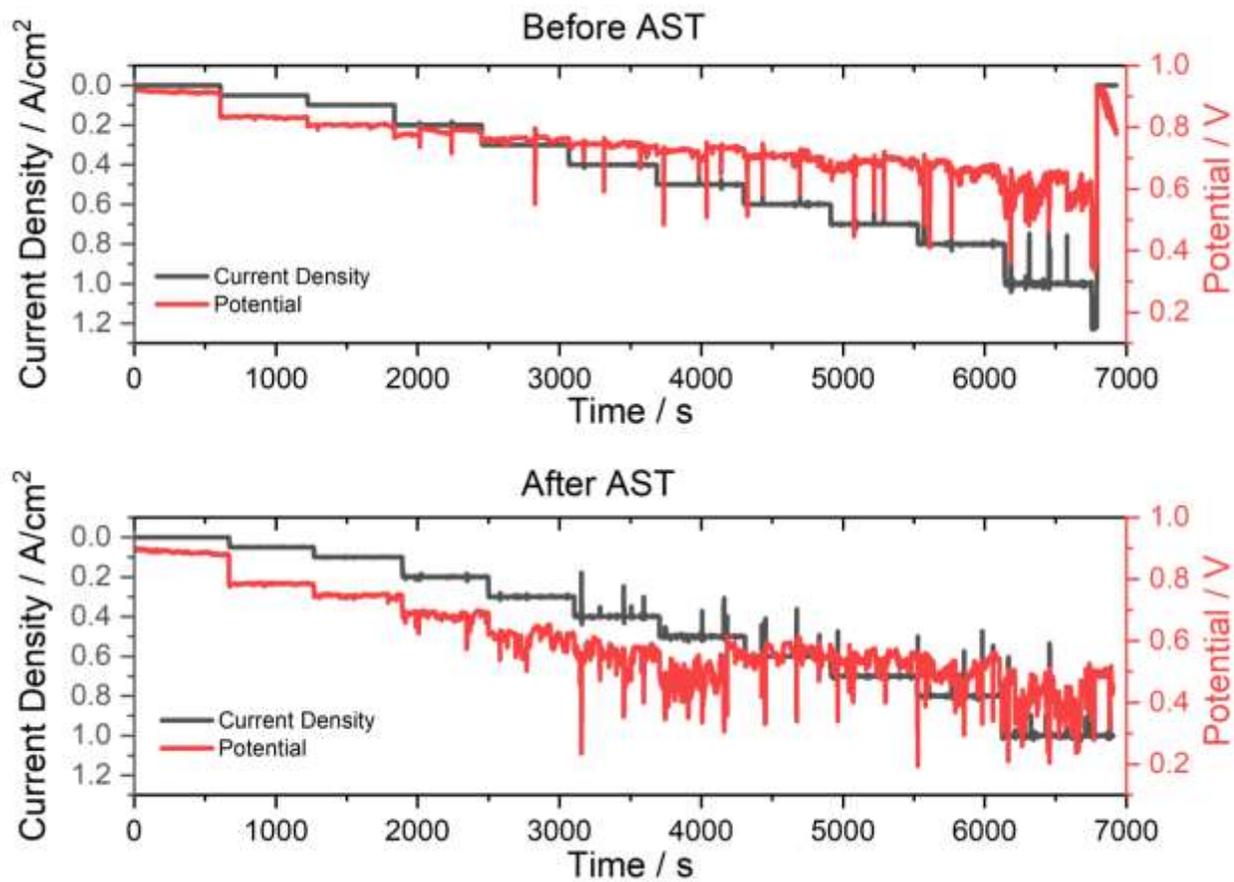

**Fig. S29.** Current-voltage measurements of MEA sample during polarization curve measurements before (top) and after (bottom) accelerated stress testing.

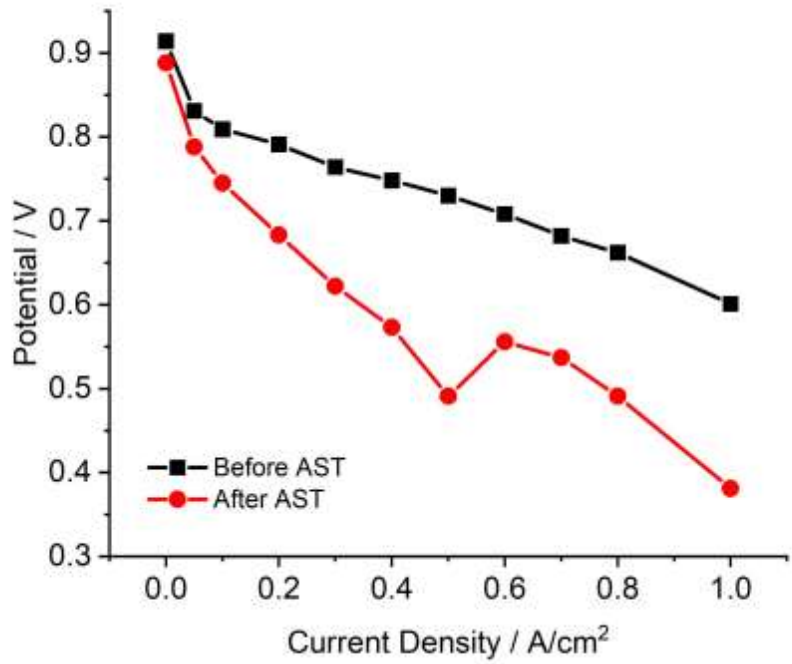

**Fig. S30.** Polarization curves of MEA sample before and after accelerated stress testing, extracted from Fig. S28.

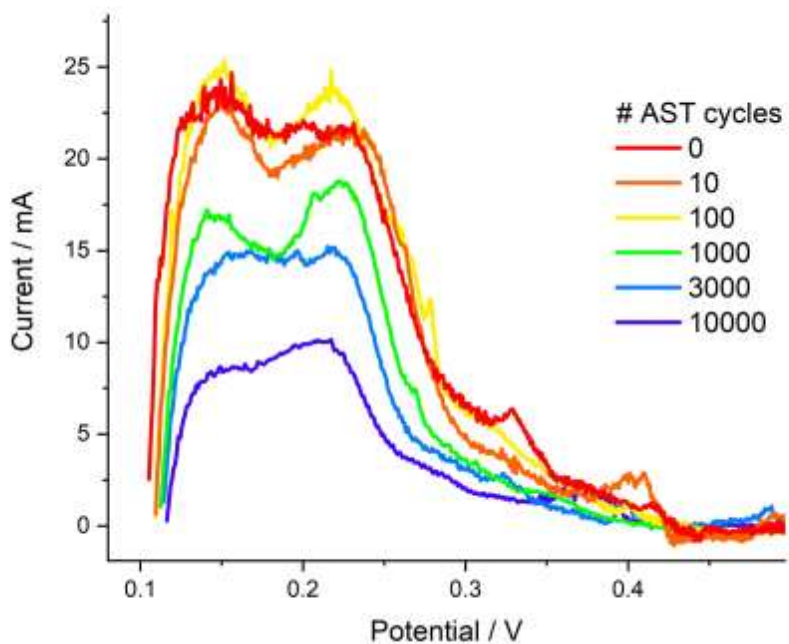

**Fig. S31.** Cyclic voltammogram of the hydrogen desorption region for the MEA catalyst, after iR correction and capacitance subtraction. This region was integrated to calculate the electrochemical surface area of the cathode catalyst. The area under each curve represents the electrochemical surface area of Pt remaining after the specified number of accelerated stress test cycles.

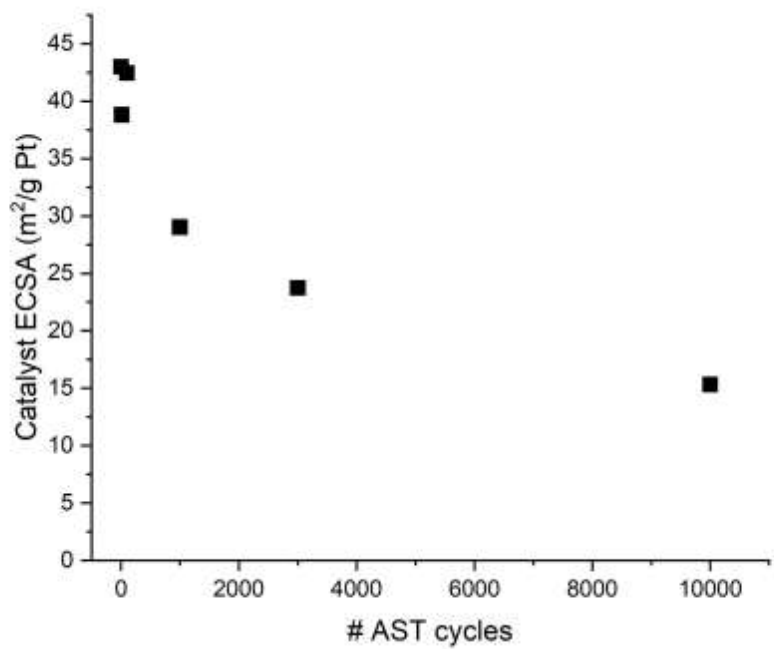

**Fig. S3222.** Electrochemical surface area of the MEA cathode catalyst during accelerated stress testing, determined using coulometry of the hydrogen desorption region (Fig. S30)

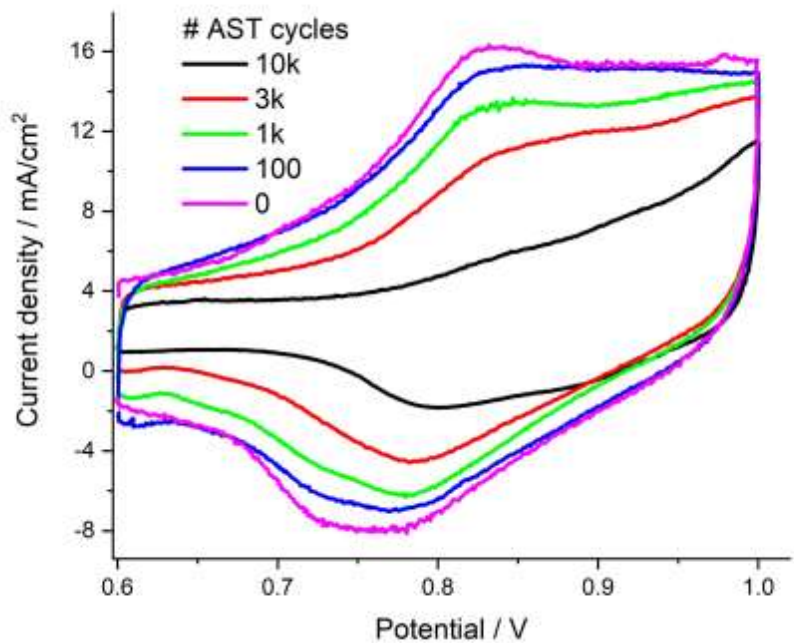

**Fig. S33.** Cyclic voltammograms collected during accelerated stress testing at selected numbers of cycles.

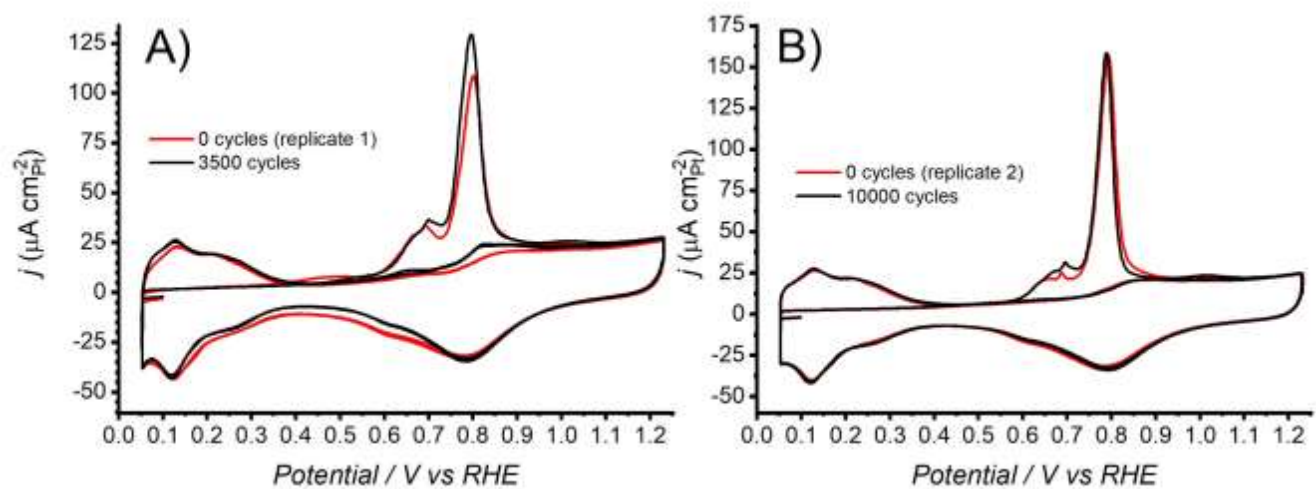

**Fig. S34.** CO stripping voltammetry of the RDE catalyst before and after accelerated stress testing.

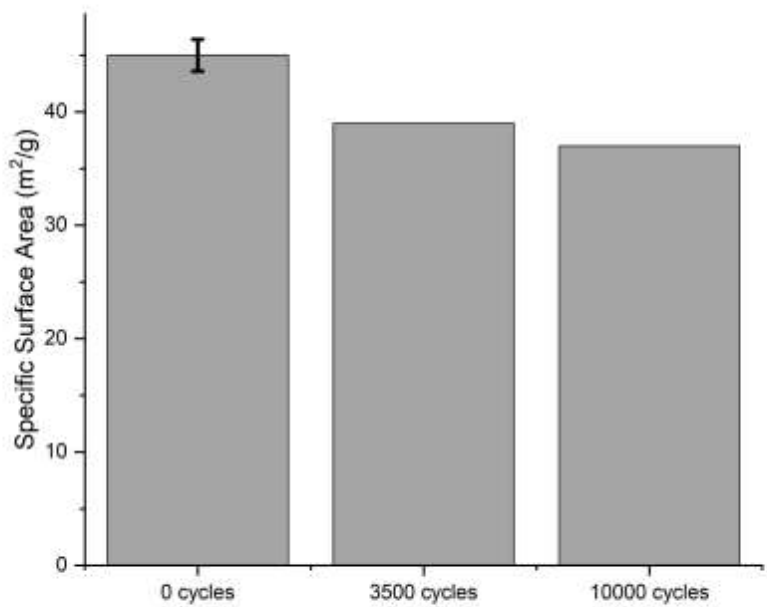

**Fig. S35.** Electrochemical surface area of the RDE catalyst before, and after 3500, and 10000 cycles of accelerated stress testing, determined using CO stripping voltammetry. Error bars correspond to the standard deviation of the two replicates.

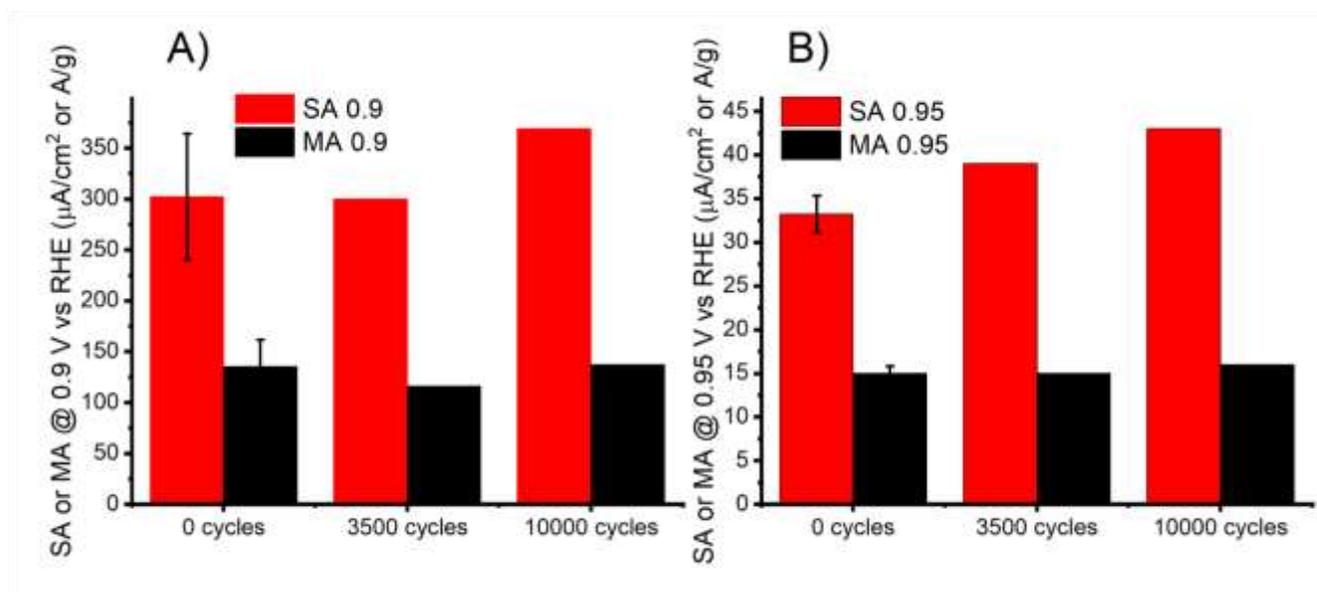

**Fig. S36.** Specific activity (SA, μA/cm$^2$) and mass activity (MA, A/g) of the RDE catalyst for the ORR before and after accelerated stress testing. The ORR activities at 0.9 V and 0.95 V vs RHE are reported in A) and B) respectively.